\documentclass{sig-alternate-05-2015}
\usepackage{hyperref}
\usepackage[T1, OT1]{fontenc}
\renewcommand{\dh}{\fontencoding{T1}\selectfont{\symbol{240}}}
\usepackage{lmodern}
\usepackage{graphicx}
\usepackage{multirow}
\setcopyright{none}

\begin{document}
\title{Comparing Published Scientific Journal Articles\\to Their Pre-Print Versions}
\subtitle{-- Extended Version --}
\numberofauthors{4}
\author{
%
% 1st. author
\alignauthor
Martin Klein \\
        \affaddr{Los Alamos National Laboratory}\\
        \affaddr{Los Alamos, NM, USA}\\
        \affaddr{\url{http://orcid.org/0000-0003-0130-2097}}\\
        \email{mklein@lanl.gov}
%\and
\alignauthor
Peter Broadwell \\
        \affaddr{University of California Los Angeles}\\
        \affaddr{Los Angeles, CA, USA}\\
        \affaddr{\url{http://orcid.org/0000-0003-4371-9472}}\\
        \email{broadwell@library.ucla.edu}
\and
\alignauthor
Sharon E. Farb \\
        \affaddr{University of California Los Angeles}\\
        \affaddr{Los Angeles, CA, USA}\\
        \affaddr{\url{http://orcid.org/0000-0002-7655-1971}}\\
        \email{farb@library.ucla.edu}
%\and
\alignauthor
Todd Grappone\\
        \affaddr{University of California Los Angeles}\\
        \affaddr{Los Angeles, CA, USA}\\
        \affaddr{\url{http://orcid.org/0000-0003-2218-7200}}\\
        \email{grappone@library.ucla.edu}
}
\maketitle
\begin{abstract}
Academic publishers claim that they add value to scholarly communications by coordinating reviews and 
contributing and enhancing text during publication. These contributions come at a considerable cost: 
U.S. academic libraries paid $\$1.7$ billion for serial subscriptions in $2008$ alone.
Library budgets, in contrast, are flat and not able to keep pace with serial price 
inflation. We have investigated the publishers' value proposition by conducting a comparative study of 
%pre-print papers from two distinct corpora and their final published counterparts. This comparison had 
pre-print papers from two distinct 
science, technology, and medicine (STM) 
corpora and their final published counterparts. This comparison had 
two working assumptions: 
1) if the publishers' argument is valid, the text of a pre-print paper should vary measurably from 
its corresponding final published version, and 2) by applying standard similarity measures, we should 
be able to detect and quantify such differences. 
Our analysis revealed that the text contents of the scientific papers generally changed very little 
from their pre-print to final published versions. These findings contribute empirical indicators 
to discussions of the added value of commercial publishers and therefore should influence libraries' 
economic decisions regarding access to scholarly publications.  
\end{abstract}
\section{Introduction}
Academic publishers of all types claim that they add value to scholarly communications by coordinating 
reviews and contributing and enhancing text during publication. These contributions come at a 
considerable cost: U.S. academic libraries paid $\$1.7$ billion for serial subscriptions in $2008$ 
alone and this number continues to rise. Library budgets, in contrast, are flat and not able to keep 
pace with serial price inflation. Several institutions have therefore discontinued or significantly 
scaled back their subscription agreements with commercial publishers such as Elsevier and 
Wiley-Blackwell. %At the University of California, Los Angeles (UCLA), 
We have investigated the publishers' value proposition by conducting a comparative study of pre-print 
papers and their final published counterparts in the areas of science, technology, and medicine (STM). 
We have two working assumptions: 
\begin{enumerate}
\item If the publishers' argument is valid, the text of a pre-print paper should vary measurably from 
its corresponding final published version. 
\item By applying standard similarity measures, we should be able to detect and quantify such 
differences. 
\end{enumerate}
In this paper we present our preliminary results based on pre-print publications from arXiv.org and 
bioRxiv.org and their final published counterparts. After matching papers via their digital object 
identifier (DOI), we applied comparative analytics and evaluated the textual similarities of 
components of the papers such as the title, abstract, and body. Our analysis revealed that the text
of the papers in our test data set changed very little from their pre-print to final published versions, 
although more copyediting changes were evident in the paper sets from bioRxiv.org than those from 
arXiv.org. In general, our results suggest that the contents of the vast majority of final published 
papers are largely indistinguishable from their pre-print versions. This work contributes empirical 
indicators to discussions of the value that academic publishers add to scholarly communication 
and therefore can influence libraries' economic decisions regarding access to scholarly publications.
\section{Global Trends in Scientific and Scholarly Publishing}
There are several global trends that are relevant and situate the focus of this research. The first 
is the steady rise in both cost and scope of the global STM 
publishing market. According to Michael Mabe and Mark Ware in their STM Report 
$2015$ \cite{ware:stmreport}, the global STM market in $2013$ was $\$25.2$ billion annually, 
with $40\%$ of this from journals ($\$10$ billion) and $68\%-75\%$ coming directly out of library 
budgets. Other relevant trends are the growing global research corpus \cite{bornmann:2015growth}, 
the steady rise in research funding \cite{ucla:accountability}, and the corresponding recent increase 
in open access publishing \cite{bjork:megajournals}. One longstanding yet infrequently mentioned 
factor is the critical contribution of faculty and researchers to the creation and establishment 
of journal content that is then licensed back to libraries to serve students, faculty 
and researchers. For example, a $2015$ Elsevier study (reported in \cite{ucla:accountability}) 
conducted for the University of California (UC) system showed that UC research publications 
accounted for $8.3\%$ of all research publications in the United States between $2009$ and 
$2013$ and the UC libraries purchased all of that research back from Elsevier.  
\subsection{The Price of Knowledge}
While there are many facets to the costs of knowledge, the pricing of published scholarly 
literature is one primary component. Prices set by publishers are meant to maximize profit and 
therefore are determined not by actual costs, but by what the market will bear. According to the 
National Association of State Budget Officers, $24$ states in the U.S. had budgets in $2013$ with 
lower general fund expenditures in $FY13$ than just prior to the global recession in 
$2008$ \cite{budget:fy14}. Nearly half of the states therefore had not returned to pre-recession 
levels of revenue and spending.
\subsection{Rise in Open Access Publications }
Over the last several years there has been a significant increase in open access publishing and 
publications in STM. Some of this increase can be traced to recent U.S. federal guidelines and 
other funder policies that require open access publication. Examples include such policies at the 
National Institutes of Health, the Wellcome Trust, and the Howard Hughes Medical Center. 
Bo-Christer Bj{\"o}rk et al. \cite{bjork:openaccess} found that in $2009$, approximately $25\%$ 
of science papers were open access. By $2015$, another study by Hammid R. Jamali and Maijid 
Nabavi \cite{jamali:openaccess} found that $61.1\%$ of journal articles were freely available 
online via open access. 
\subsection{Pre-Print versus Final Published Versions and the Role of Publishers}
In this study, we compared paper pre-prints from the arXiv.org and bioRxiv.org repositories to 
the corresponding final published versions of the papers. The annual budget for arXiv.org as posted on
the repository's wiki is set at an average of $\$826,000$ for 
$2013-2017$.\footnote{\url{https://confluence.cornell.edu/display/arxivpub/arXiv+Public+Wiki}} While we do 
not have access to the data to precisely determine the corresponding costs for commercial publishing, the 
National Center for Education Statistics found in $2013$ that the market for English-language STM journals 
was approximately $\$10$ billion annually. It therefore seems safe to say that the costs for commercial 
publishing are orders of magnitude larger than the costs for organizations such as arXiv.org and 
bioRxiv.org.

Michael Mabe describes the publishers' various roles as including, but not limited 
to entrepreneurship, copyediting, tagging, marketing, distribution, and e-hosting \cite{mabe:ermh}.
The focus of the study presented here is on the publishers' contributions to the content of the materials 
they publish (specifically copyediting and other enhancements to the text) and how and to what extent, if 
at all, the content changes from the pre-print to the final published version of a publication.
This article does not consider other roles publishers play, for example, with respect to entrepreneurship, 
tagging, marketing, distributing, and hosting.
\section{Data Gathering}
Comparing pre-prints to final published versions of a significant corpus of scholarly articles from
science, technology, and medicine required obtaining the contents of both versions of each article in 
a format that could be analyzed as full text and parsed into component sections (title, abstract, body) 
for more detailed comparisons.  The most accessible sources of such materials proved to be 
\texttt{arXiv.org} and \texttt{bioRxiv.org}.

arXiv.org is an open access digital repository owned and operated by Cornell University and supported by 
a consortium of institutions. At the time of writing, arXiv.org hosts over $1.2$ million academic 
pre-prints, most written in fields of physics and mathematics and uploaded by their authors to the site 
within the past $20$ years. The scope of arXiv.org enabled us to identify and obtain a sufficiently 
large comparison corpus of corresponding final published versions in scholarly journals to which our 
institution has access via subscription.

bioRxiv.org is an open access repository devoted specifically to unrefereed pre-prints (papers that have 
not yet been peer-reviewed for publication) in the life sciences, operated by Cold Spring Harbor 
Laboratory, a private, nonprofit research institution. It began accepting papers in late $2013$ and
at the time of writing hosts slightly more than $10,000$ pre-prints. bioRxiv is thus much smaller than 
arXiv, and most of the corresponding final published versions in our bioRxiv data set were obtained via 
open access publications, rather than those accessible only via institutional subscriptions. Nonetheless,
because bioRxiv focuses on a different range of scientific disciplines and thus archives pre-prints of 
papers published in a largely distinct set of journals, an analysis using this repository provides an
informative contrast to our study of arXiv.

\subsection{arXiv Corpus}
Gathering pre-print texts from arXiv.org proceeded via established public interfaces for machine 
access to the site data, respecting their discouragement of indiscriminate automated 
downloads.\footnote{\url{https://arxiv.org/help/robots}}

We first downloaded metadata records for all articles available from arXiv.org through February 
of $2015$ via the site's Open Archives Initiatives Protocol for Metadata Harvesting (OAI-PMH) 
interface\footnote{\url{http://export.arxiv.org/oai2?verb=Identify}}.
We received $1,015,440$ records in all, which provided standard Dublin Core metadata for each article, 
including its title and authors, as well as other useful data for subsequent analysis, such as the 
paper's disciplinary category within arXiv.org and the upload dates of its versions (if the authors 
submitted more than one version). The metadata also contained the text of the abstract for most 
articles. Because the abstracts as well as the article titles often contained text formatting 
markup, however, we preferred to use instances of these texts that we derived from other sources, 
such as the PDF version of the paper, for comparison purposes (see below).

arXiv.org's OAI-PMH metadata record for each article contains a field for a DOI, which we used as 
the key to match pre-print versions of articles to their final published versions. arXiv.org does 
not require DOIs for submitted papers, but authors may provide them voluntarily. $452,017$ article 
records in our initial metadata set ($44.5\%$) contained a DOI. Working under the assumption that 
the DOIs are correct and sufficient to identify the final published version of each article, we 
then queried the publisher-supported CrossRef citation linking 
service\footnote{\url{https://github.com/CrossRef/rest-api-doc/blob/master/rest_api.md}} 
to determine whether the full text of the corresponding published article would be available for 
download via UCLA's institutional journal subscriptions.

To begin accumulating full articles for text comparison, we downloaded PDFs of every pre-print 
article from arXiv.org with a DOI that could be matched to a full-text published version
accessible through subscriptions held by the UCLA Library.
Our initial query indicated that up to $12,666$ final published versions would be accessible in 
this manner. 
The main reason why this number is fairly low is that, at the time of writing, the above mentioned 
CrossRef API is still in its early stages and only a few publishers have agreed to making their 
articles available for text and data mining via the API. 
However, while this represented a very small proportion of all papers with DOI-associated pre-prints 
stored in arXiv.org ($2.8$\% at the time of the analysis), the resulting collection nevertheless was 
sufficient for a detailed comparative analysis. Statistically, a random sample of this size would be
more than adequate to provide a $95$\% confidence level; our selection of papers was not truly
random, but as noted below, the similar proportions of paper subject areas in our corpus to the 
proportions of subject areas among all pre-prints in arXiv.org also provides a positive indicator of 
its representativeness.

The downloads of pre-prints took place via arXiv.org's bulk data access service, which facilitates 
the transfer of large numbers of articles as PDFs or as text markup source files and images, 
packaged into .tar archives, from an Amazon S3 account. Bandwidth fees are paid by the requesting 
party.\footnote{\url{https://arxiv.org/help/bulk_data_s3}}
This approach only yields the most 
recent uploaded version of each pre-print article, however, so for analyses involving earlier 
uploaded versions of pre-print articles, we relied upon targeted downloads of earlier article 
versions via arXiv.org's public web interface.
\subsection{arXiv Corpus of Matched Articles}
Obtaining the final published versions of article pre-prints from arXiv.org involved querying the 
CrossRef API to find a full-text download URL for a given DOI. Most of the downloaded files 
($96$\%) arrived in one of a few standard XML markup formats; the rest were in PDF format. Due to 
missing or incomplete target files, $464$ of the downloads failed entirely, leaving us with 
$12,202$ published versions for comparison. The markup of the XML files contained, in addition to 
the full text, metadata entries from the publisher. Examination of this data revealed that the 
vast majority ($99\%$) of articles were published between $2003$ and $2015$. This time range 
intuitively makes sense as DOIs did not find widespread adoption with commercial publishers until 
the early $2000s$. 
\begin{figure}[t!]
\center
\includegraphics[scale=0.17]{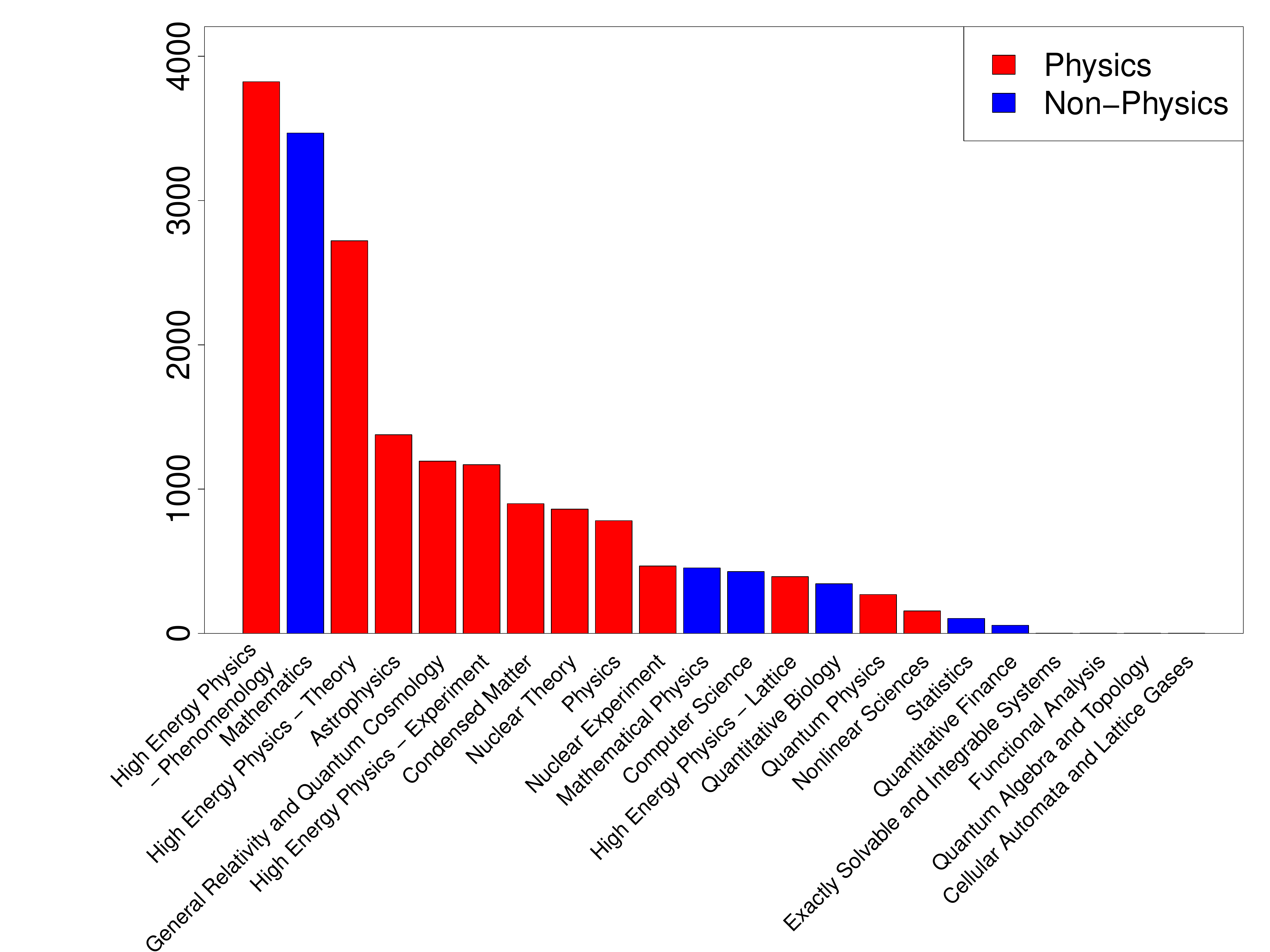}
\caption{arXiv.org categories of matched articles}
\label{fig:arxiv_categories}
\end{figure}

The disciplines of articles in arXiv.org are dominated by physics, mathematics, statistics, and 
computer science.\footnote{\url{https://arxiv.org/help/stats/2016_by_area/index}} We found a 
very similar distribution of categories in our corpus of matched articles, as shown in 
Figure \ref{fig:arxiv_categories}.
An overview of the journals in which the matched articles are published is provided in the left
half of Table \ref{tab:journal_overview}. The data shows that most of the obtained published versions 
($96$\%) were published in Elsevier journals.
\subsection{arXiv Corpus Data Preparation}
For this study, we compared the texts of the titles, abstracts, and body sections of the pre-print 
and final published version of each paper in our data set. Being able to generate these sections 
for most downloaded papers therefore was a precondition of this analysis.

All of the pre-print versions and a small minority of final published papers ($4\%$)
were downloaded in PDF format. To identify and extract the sections of these papers, we used the 
GROBID\footnote{\url{https://github.com/kermitt2/grobid}} 
library, which employs trained conditional random field machine learning algorithms to segment 
structured scholarly texts, including article PDFs, into XML-encoded text.

The markup tags of the final published papers downloaded in XML format usually identified quite 
plainly their primary sections. A sizable proportion ($11\%$) of such papers, however, did not contain a 
demarcated body section in the XML and instead only provided the full text of the papers. Although it is 
possible to segment these texts further via automatic scholarly information extraction tools such as 
ParsCit,\footnote{\url{http://aye.comp.nus.edu.sg/parsCit/}} which use trained conditional random field 
models to detect sections probabilistically, for the present study we elected simply to omit the body 
sections of this subset of papers from the comparison analysis.
\begin{figure}[t!]
\center
\includegraphics[scale=0.17]{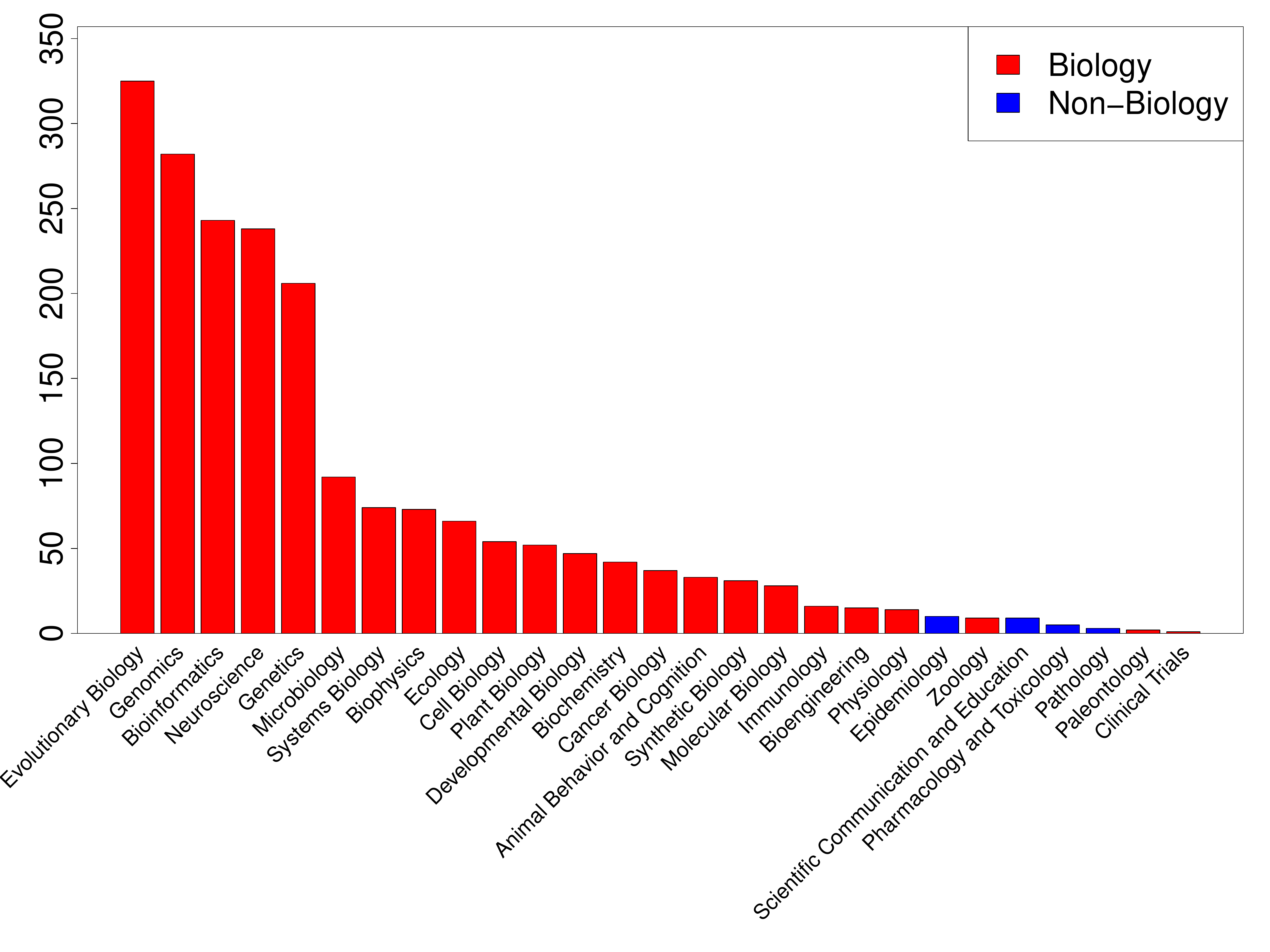}
\caption{bioRxiv.org subjects of matched articles}
\label{fig:biorxiv_subjects}
\end{figure}

As noted above, the GROBID software used to segment the PDF papers was probabilistic in its approach, 
and although it was generally quite effective, it was not able to isolate all sections (title, 
abstract, body) for approximately $10-20$\% of the papers in our data set. This situation, combined 
with the aforementioned irregularities in the XML of a similar proportion of final published papers, 
meant that the number of corresponding texts for comparison varied by section. Thus, 
for our primary comparison of the latest pre-print version uploaded to arXiv.org to its final 
published version, we were able to compare directly $10,900$ titles and abstract sections and 
$9,399$ body sections.

The large variations in formatting of the references sections (also called the ``tail'') as extracted 
from the raw downloaded XML and the parsed PDFs, however, precluded a systematic comparison of that 
section. We leave such an analysis for future work. A further consequence of our text-only analysis 
was that the contents of images were ignored entirely, although figure captions and the text contents 
of tables usually could be compared effectively.
\begin{table*}[t!]
\center
\caption{Overview of top 20 journals of final published versions per corpus}
\begin{tabular}{|c|l||c|l|} \hline
\multicolumn{2}{|c||}{\textbf{arXiv Corpus}} & \multicolumn{2}{c|}{\textbf{bioRxiv Corpus}} \\ 
Freq & Journal & Freq & Journal \\ \hline \hline
7143 & Physics Letters B & 154 & PLOS ONE \\ \hline
261 & Journal of Algebra & 98 & Scientific Reports \\ \hline
229 & Nuclear Physics B & 91 & Genetics \\ \hline
218 & Advances in Mathematics & 86 & eLife \\ \hline
179 & Biophysical Journal & 69 & PLOS Genetics \\ \hline
& \multirow{1}{*}{Nuclear Instruments and Methods in Physics Research Section A:} & & \\
179 & Accelerators, Spectrometers, Detectors and Associated Equipment & 69 & PLOS Computational Biology \\ \hline
175 & Physics Letters A & 66 & PNAS \\ \hline
162 & Journal of Mathematical Analysis and Applications & 59 & G3: Genes|Genomes|Genetics \\ \hline
154 & Physica A: Statistical Mechanics and its Applications & 52 & Genome Biology \\ \hline
146 & Journal of Functional Analysis & 46 & Nature Communications \\ \hline
125 & Annals of Physics & 44 & BMC Genomics \\ \hline
122 & Linear Algebra and its Applications & 42 & Genome Research \\ \hline
122 & Nuclear Physics A & 33 & BMC Bioinformatics \\ \hline
107 & Computer Physics Communications & 26 & Molecular Ecology \\ \hline
104 & Journal of Pure and Applied Algebra & 26 & Nature Genetics \\ \hline
96 & Topology and its Applications & 25 & NeuroImage \\ \hline
96 & Journal of Number Theory & 24 & PeerJ \\ \hline
80 & Theoretical Computer Science & 23 & Evolution \\ \hline
77 & Stochastic Processes and their Applications & 19 & Nature Methods \\ \hline
73 & Icarus & 19 & American Journal of Human Genetics \\ \hline
\end{tabular}
\label{tab:journal_overview}
\end{table*}
\subsection{bioRxiv Corpus}
Compared to the arXiv papers, we were able to accumulate a smaller but overall more proportionately 
representative corpus of life science pre-prints and final published papers from bioRxiv.org. The 
repository does not as yet offer the same sophisticated bulk metadata access and PDF downloading 
features as arXiv.org, but fortunately the comparatively small scale of bioRxiv enabled us to collect 
article metadata and texts utilizing basic scripting tools. We first gathered metadata via the 
bioRxiv.org site's search and browse features for all articles posted to the site from its inception in 
November 2013 until November 2016. For these articles, which numbered $7,445$ in total, we extracted 
the author-supplied DOIs and journal information about their eventual publication venues, when provided, 
as well as titles, abstracts, download links, and submission dates for all versions of the pre-prints.
\subsection{bioRxiv Corpus of Matched Articles}
$2,516$ of the pre-print records in bioRxiv contained final publication DOIs. We attempted to obtain the
full texts of the published versions by querying these DOIs via the CrossRef API as described above for
the arXiv papers. Relatively few of these papers --- $220$ in all --- were actually available in full
text via this method. We then used the R 'fulltext' package from the rOpenSci 
project,\footnote{\url{https://github.com/ropensci/fulltext}} which also searches sources including PLOS,
Biomed Central, and PMC/Pubmed, and ultimately had more success, obtaining a total of $1,443$ published
papers with full texts and an additional $1,054$ publication records containing titles and abstracts but
no body texts or end matter sections. Most of the primary subjects of these matched articles are in the 
field of biology. The corresponding overview of subject areas is provided in Figure \ref{fig:biorxiv_subjects}. 
The journals in which the articles are published are provided in the right half of 
Table \ref{tab:journal_overview}.
\subsection{bioRxiv Corpus Data Preparation}
Extraction of the data from the bioRxiv pre-print and published articles for the text comparison
proceeded via a similar process to that of the arXiv data preparation: the earliest and latest
versions of the matched pre-print articles (as well as a handful of final published papers
only available as PDF) were downloaded as PDFs and parsed into their component sections via the 
GROBID software. The downloaded records of the final published versions were already separated
into these sections via XML markup, so rudimentary parsing routines were sufficient 
to extract the texts from these files. We also extracted publication dates from these records to
facilitate the timeline analyses shown below. %in Figure \ref{fig:biorxiv_pub_dates_versions}.
\section{Analytical Methods} \label{subsec:txt_comp_meth}
We applied several text comparison algorithms to the corresponding sections of the pre-print and 
final published versions of papers in our test data set. These algorithms, described in detail below, 
were selected to quantify different notions of ``similarity'' between texts. We 
normalized the output values of each algorithm to lie between $1$ and $0$, with $1$ indicating that 
the texts were effectively identical, and $0$ indicating complete dissimilarity. Different algorithms 
necessarily measured any apparent degree of dissimilarity in different ways, so the outputs of the 
algorithms cannot be compared directly, but it is nonetheless valid to interpret the aggregation 
of these results as a general indication of the overall degree of similarity between two texts 
along several different axes of comparison.
\subsection{Editorial Changes}
The well-known Levenshtein edit distance metric \cite{levenshtein:edit_distance}
calculates the number of character insertions, deletions, and substitutions necessary to convert one 
text into another. It thus provides a useful quantification of the amount of editorial intervention 
--- performed either by the authors or the journal editors --- that occurs between the pre-print and 
final published version of a paper. Our work used the edit ratio calculation as provided in the 
Levenshtein Python C Implementation 
Module,\footnote{\url{https://pypi.python.org/pypi/python-Levenshtein/0.11.2}}
which subtracts the 
edit distance between the two documents from their combined length in characters and divides this 
amount by their aggregate length, thereby producing a value between $1$ (completely similar) and 
$0$ (completely dissimilar).
\subsection{Length Similarity}
The degree to which the final published version of a paper is shorter or longer than the pre-print 
constitutes a much less involved but nonetheless revealing comparison metric. To calculate this 
value, we divided the absolute difference in length between both papers by the length of the longer 
paper and subtracted this value from $1$. Therefore, two papers of the same length will receive a 
similarity score of $1$; this similarity score is $0.5$ if one paper is twice as long as the other, 
and so on. It is also possible to incorporate the polarity of this change by adding the length 
ratio to $0$ if the final version is longer, and subtracting it from $0$ if the pre-print is longer.
\begin{figure*}[t!]
\center
\includegraphics[scale=0.5]{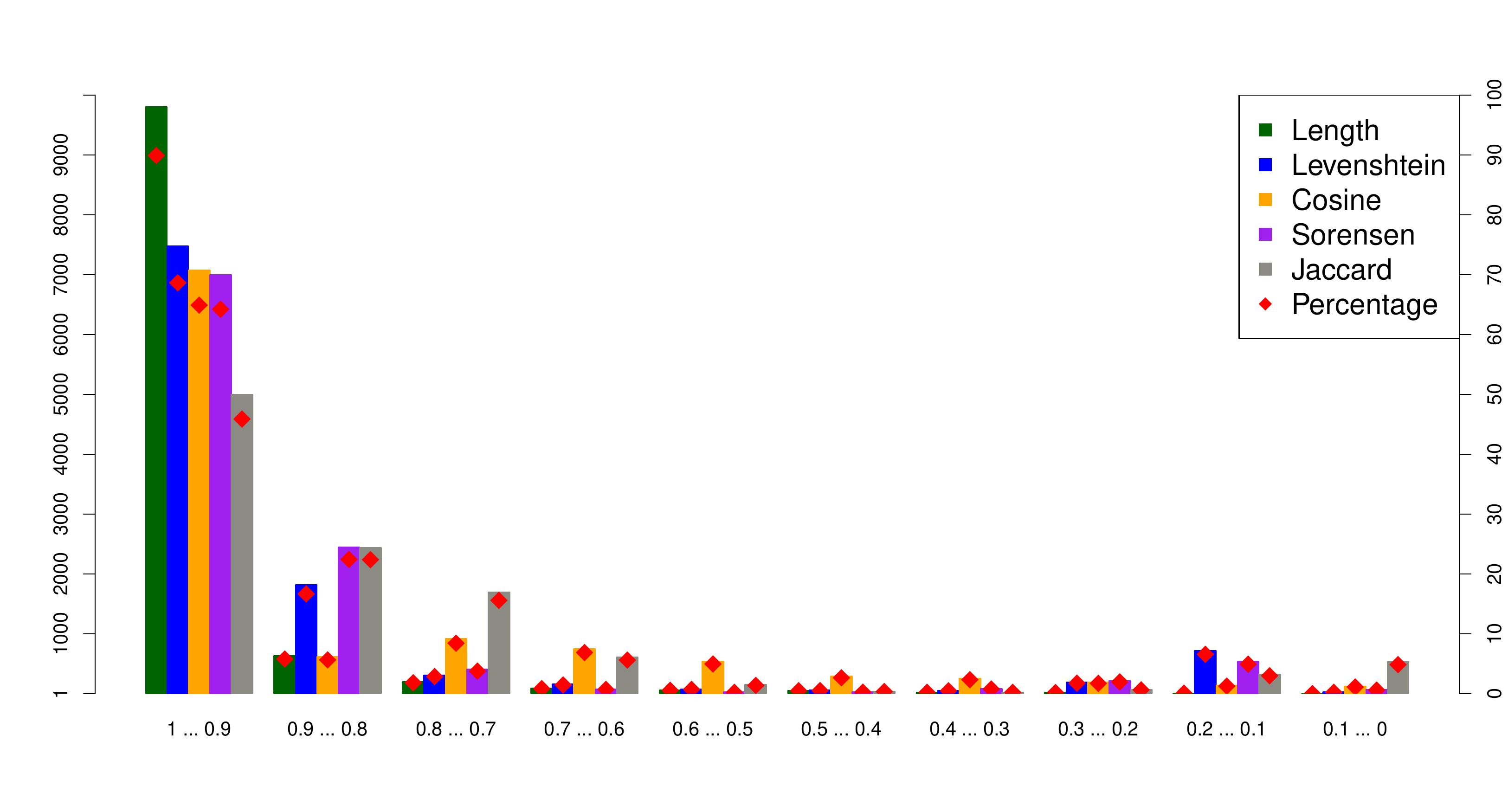}
\caption{arXiv corpus - comparison results for titles}
\label{fig:title_histo}
\end{figure*}
\subsection{String Similarity}
Two other fairly straightforward, low-level metrics of string similarity that we applied to the paper 
comparisons were the Jaccard and S\o rensen indices, which consider only the sets of unique characters 
that appear in each text. The S\o rensen similarity \cite{sorensen:index} was calculated by doubling 
the number of unique characters shared between both texts (the intersection) and dividing this by 
the combined sizes of both texts' unique character sets. 

The Jaccard similarity calculation \cite{jaccard:index} is the size of the intersection (see above) 
divided by the total number of unique characters appearing in either the pre-print or final published 
version (the union). 

Implementations of both algorithms were provided by the standard Python string distance 
package.\footnote{\url{https://pypi.python.org/pypi/Distance/}}
\subsection{Semantic Similarity}
Comparing overall lengths, shared character sets, and even edit distances between texts does not 
necessarily indicate the degree to which the meaning of the texts --- that is, their semantic 
content --- actually has changed from one version to another. To estimate this admittedly more 
subjective notion of similarity, we calculated the pairwise cosine similarity between the pre-print 
and final published texts. Cosine similarity can be described intuitively as a measurement of how 
often significant words occur in similar quantities in both texts, normalized by the lengths of 
both documents \cite{pang:introdm}. The actual procedure used for this study involved removing 
common English ``stopwords'' from each document, then applying the Porter stemming 
algorithm \cite{porter:algo} to remove suffixes and thereby merge closely related words, before 
finally applying the pairwise cosine similarity algorithm implemented in the Python scikit-learn 
machine learning package\footnote{\url{http://scikit-learn.org/stable/}} to the resulting term 
frequency lists. Because this implementation calculates only the similarity between two documents 
considered in isolation, instead of within the context of a larger corpus, it uses raw term counts, 
rather than term-frequency/inverse document frequency (TF-IDF) weights.
\section{arXiv Corpus\\Experiment Results} \label{sec:arxiv_experiments}
We calculated the similarity metrics described above for each pair of corresponding pre-print and 
final published papers in our data set from arXiv.org, comparing titles, abstracts, and body sections.
See Section \ref{sec:biorxiv_results} for the results of running the same comparisons on the 
papers from bioRxiv.org. From the results of these calculations, we generated visualizations of the 
similarity distributions for each metric. Subsequent examinations and analyses of these distributions 
provided novel insights into the question of how and to what degree the text contents of scientific 
papers may change from their pre-print instantiations to the final published version. Because each 
section of a publication differs in its purpose and characteristics (e.g., length, standard 
formatting) and each metric addresses the notion of similarity from a different perspective, we 
present the results of our comparisons per section (title, abstract, and body), subdivided by 
comparison metric.
\begin{figure*}[ht!]
\center
\includegraphics[scale=0.5]{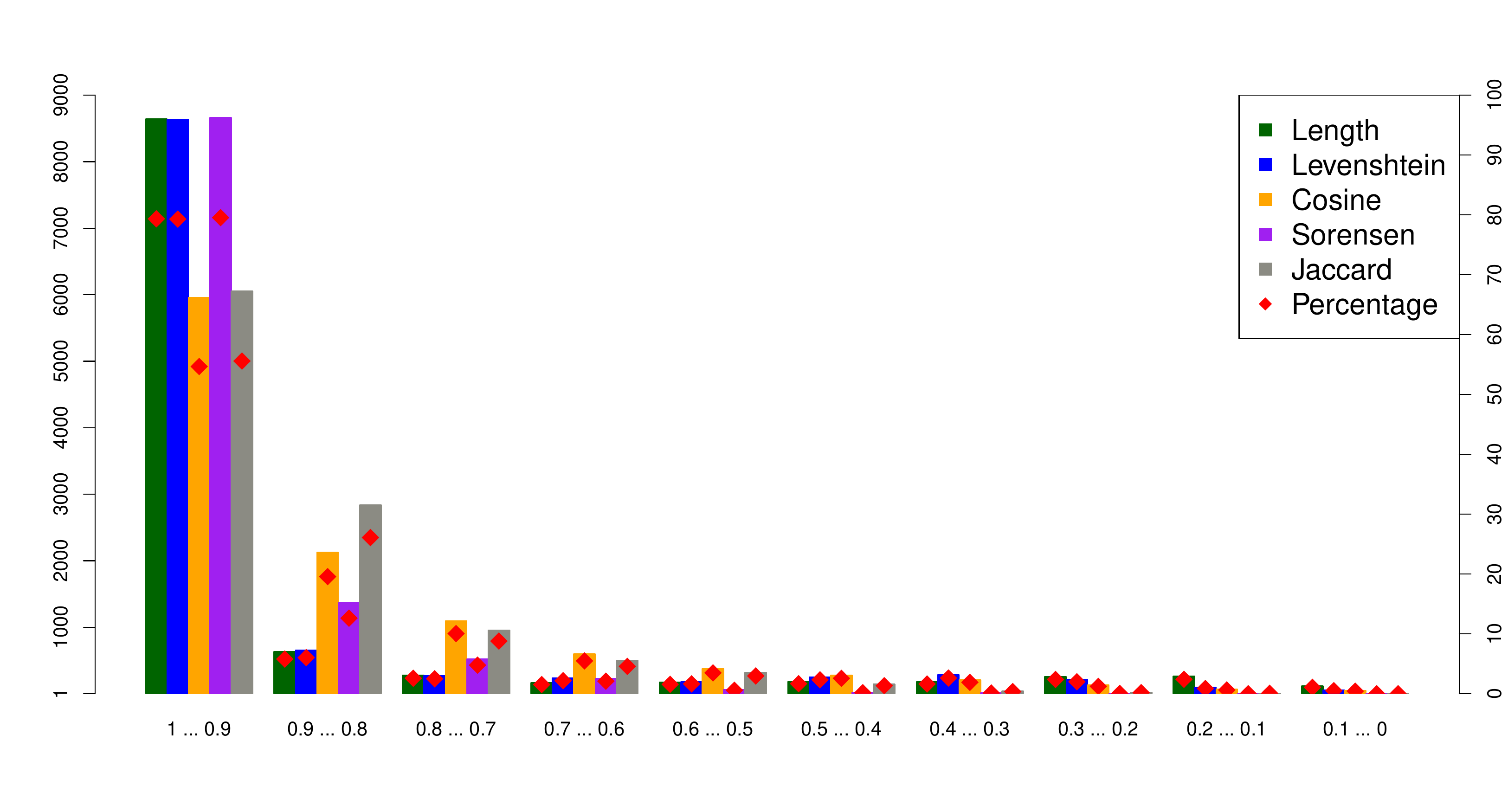}
\caption{arXiv corpus - comparison results for abstracts}
\label{fig:abstract_histo}
\end{figure*}
\subsection{Title Analysis}
First, we analyzed the papers' titles. A title is usually much shorter (fewer characters) than a 
paper's abstract and its body. That means that even small changes to the title would have a large 
impact on the similarity scores based on length ratio and Levenshtein distance. Titles also often 
contain salient keywords describing the overall topic of the paper. If those keywords were changed, 
removed or new ones added, the cosine similarity value would drop.

Figure \ref{fig:title_histo} shows the comparison of results of all five text similarity measures 
applied to titles. Since all measures are normalized, their values range between $0$ and $1$.
Values close to $1$ indicate a very high level of similarity and values close to $0$ represent a 
high degree of dissimilarity of the analyzed text. 
Figure \ref{fig:title_histo} shows results aggregated into ten bins. Each bin represents a range of
the normalized score. For example, the leftmost bin represents scores between $1.0$ and $0.9$, the
bin second to the left represents scores between $0.9$ and $0.8$, and the rightmost bin
represents scores between $0.1$ and $0$. The horizontal x-axis shows the ranges the bins represent.
Each bin contains five bars, one for each similarity measure applied. The height of a bar indicates 
the number of articles whose title similarity score falls into the corresponding bin. For example,
imagine an article that has the following title similarity scores: $Length = 0.93$, 
$Levenshtein = 0.91$, $Cosine = 0.83$, $S\o rensen = 0.75$, and $Jaccard = 0.73$. This article would 
therefore contribute to the green and dark blue bars in the leftmost bin, to the yellow bar in the bin 
second from the left, and to the purple and gray bars in the bin third from the left.
The total height of a bar (the total number of articles) can be read from the left y-axis in absolute
numbers.
In addition, each bar has a red diamond that shows its magnitude relative to the size of the entire corpus. 
This percentage can be read from the right y-axis.

Figure \ref{fig:title_histo} shows a dominance of the top bin. The vast majority of titles have a 
very high score in all applied similarity measures. Most noticeably, almost $10,000$ titles 
(around $90\%$ of all titles) are of very similar length, with a ratio value between $1$ and $0.9$.
The remaining $10\%$ fall into the next bin with values between $0.9$ and $0.8$.
A very similar observation can be made for the Levenshtein distance and the S\o rensen value. About 
$70\%$ of those values fall into the top bin and the majority of the remaining values (around $20\%$)
land between $0.9$ and $0.8$.  
The cosine similarity is also dominated by values in the top bin (around $70\%$) but the remaining 
values are more distributed across the second, third, fourth, and fifth bin.
Just about half of all Jaccard values can be seen in the top bin and most of the remainder is split 
between the second ($25\%$) and the third bin ($20\%$). In many cases, this metric is registering 
low-level but systematic differences in character use between the pre-print and final published 
versions as filtered through the download methods described above: for example, a pre-print may 
consistently use em-dashes (--), whereas the published version uses only hyphens (-). This sensitivity 
of the Jaccard similarity score to subtle changes in the unique character sets in each text is 
apparent for other sections as well.

The results of this comparison, in particular the fact that the majority of values fall between 
$1$ and $0.9$, provide very strong indicators that titles of scholarly articles do not change 
noticeably between the pre-print and the final published version. Even though 
Figure \ref{fig:title_histo} shows a small percentage of titles exhibiting a rather low level of 
similarity, with Levenshtein and S\o rensen values between $0.2$ and $0.1$, the overall similarity 
of titles is very high.
\begin{figure*}[ht!]
\center
\includegraphics[scale=0.5]{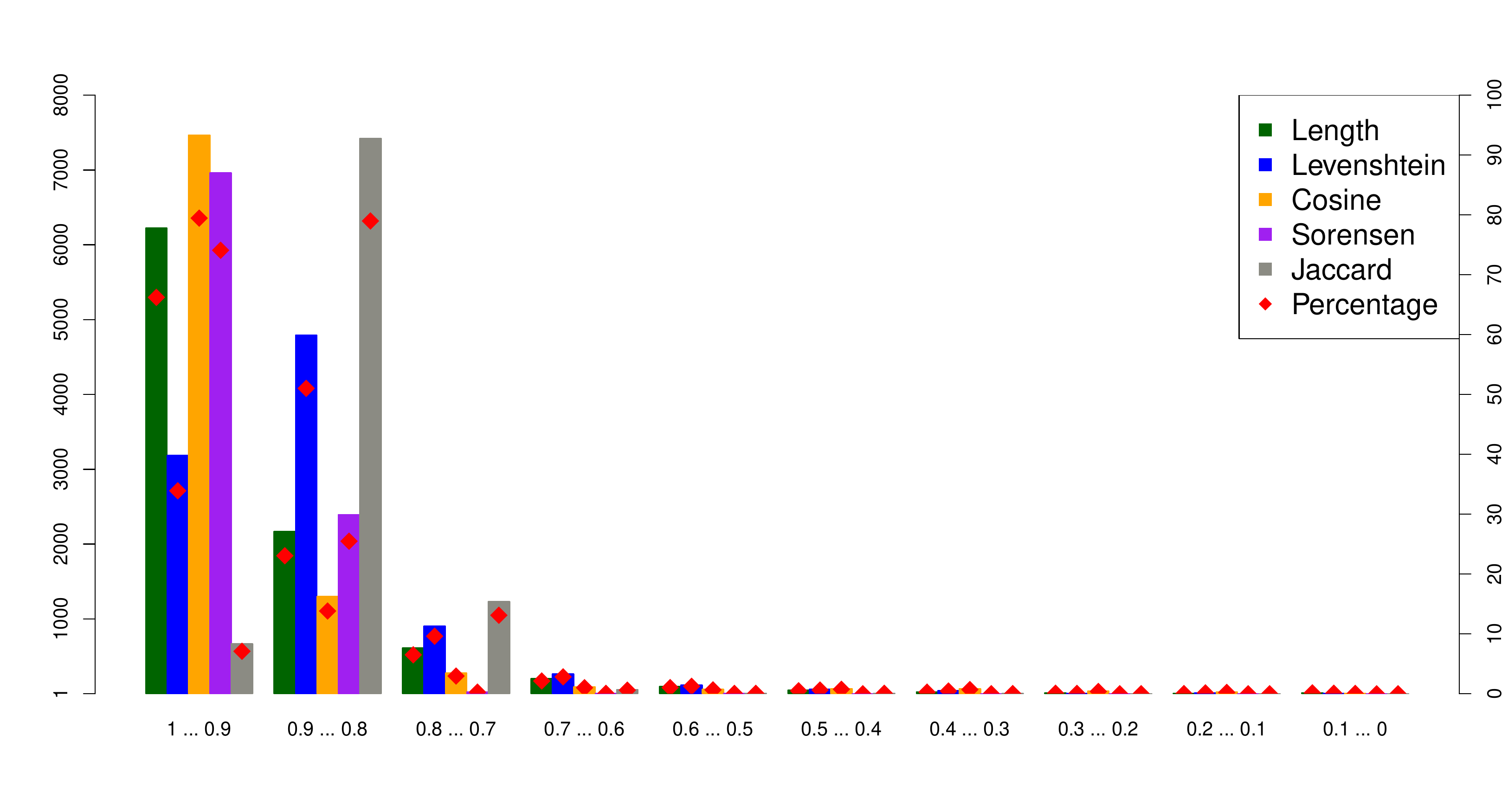}
\caption{arXiv corpus - comparison results for body sections}
\label{fig:body_histo}
\end{figure*}
\subsection{Abstract Analysis}
The next section we compared was the papers' abstracts. An abstract can be seen as a very short 
version of the paper. It often gives a brief summary of the problem statement, the methods applied, 
and the achievements of the paper. As such, an abstract usually is longer than the paper's title 
(in number of characters) and provides more context. Intuitively, it seems probable that we would 
find more editorial changes in longer sections of the pre-print version of an article compared to 
its final published version. However, a potentially increased number of editorial changes alone 
does not necessarily prove dissimilarity between longer texts. We expect similarity measures based 
on semantic features such as cosine similarity to be more reliable here.

Figure \ref{fig:abstract_histo} shows the comparative results for all abstracts. The formatting of 
the graph is the same as previously described for Figure \ref{fig:title_histo}. To our surprise, 
the figure is dominated by the high frequency of values between $1$ and $0.9$ across all similarity 
measures. More than $8,500$ abstracts (about $80\%$) have such scores for their length ratio, 
Levenshtein distance, and S\o rensen index. $6\%$ of the remaining length ratio and Levenshtein 
distance values as well as $13\%$ of the remaining S\o rensen index values fall between $0.9$ and 
$0.8$. The remaining pairs are distributed across all other bins.
The cosine similarity and Jaccard index values are slightly more distributed. About $5,000$ abstracts 
($55\%$) fall into the top bin, $20\%$ and $26\%$ into the second, and $10\%$ and $9\%$ into the 
third bin, respectively.

Not unlike our observations for titles, the algorithms applied to abstracts predominantly return 
values that indicate a very high degree of similarity. Figure \ref{fig:abstract_histo} shows that 
more than $90\%$ of abstracts score $0.6$ or higher, regardless of the text similarity measure 
applied. It is also worth pointing out that there is no noticeable increased frequency of values 
between $0.2$ and $0.1$ as previously seen when comparing titles (Figure \ref{fig:title_histo}). 
\subsection{Body Analysis} \label{sec:body_analysis}
The next section we extracted from our corpora of scholarly articles and subjected to the text 
similarity measures is the body of the text. This excludes the title, the author(s), the abstract, 
and the reference section. This section is, in terms of number of characters, the longest of our 
three analyzed sections. We therefore consider scores resulting from algorithms based on editorial 
changes to be less informative for this comparison. In particular, a finding such as ``The body 
of article $A_1$ contains $10\%$ fewer characters than the body of article $A_2$'' would not provide 
any reliable indicators of the similarity between the two articles $A_1$ and $A_2$. Algorithms based 
on semantic features, such as cosine similarity, on the other hand, provide stronger indicators 
of the similarity of the compared long texts. More specifically, cosine values are expected to be 
rather low for very dissimilar article body sections.

The results of this third comparison can be seen in Figure \ref{fig:body_histo}. The height of the 
bar representing the cosine similarity is remarkable. Almost $7,500$ body sections of our compared 
scholarly articles, which is equivalent to $80\%$ of the entire corpus, have a cosine score that 
falls in the top bin with values between $1$ and $0.9$. $14\%$ have a cosine value that falls into 
the second and $3\%$ fall into the third bin. Values of the S\o rensen index show a very similar 
pattern with $74\%$ in the top bin and $25\%$ in the second. In contrast, only $7\%$ of articles' 
bodies have Jaccard index values falling into the top bin. The vast majority of these scores, 
$79\%$, are between $0.9$ and $0.8$ and another $13\%$ are between $0.8$ and $0.7$.
It is surprising to see that even the algorithms based on editorial changes provide scores mostly 
in the top bins. Of the length ratio scores, $66\%$ fall in the top bin and $23\%$ in the second 
bin. The Levenshtein distance shows the opposite proportions: $34\%$ are in the top and $51\%$ 
belong to the second bin.

The dominance of bars on the left hand side of Figure \ref{fig:body_histo} provides yet more evidence
that pre-print articles of our corpus and their final published version do not exhibit many features 
that could distinguish them from each other, neither on the editorial nor on the semantic level. 
$95\%$ of all analyzed body sections have a similarity score of $0.7$ or higher in any of the 
applied similarity measures. 
\subsection{Publication Dates} \label{subsec:pub_dates}
\begin{figure}[t!]
\center
\includegraphics[scale=0.17]{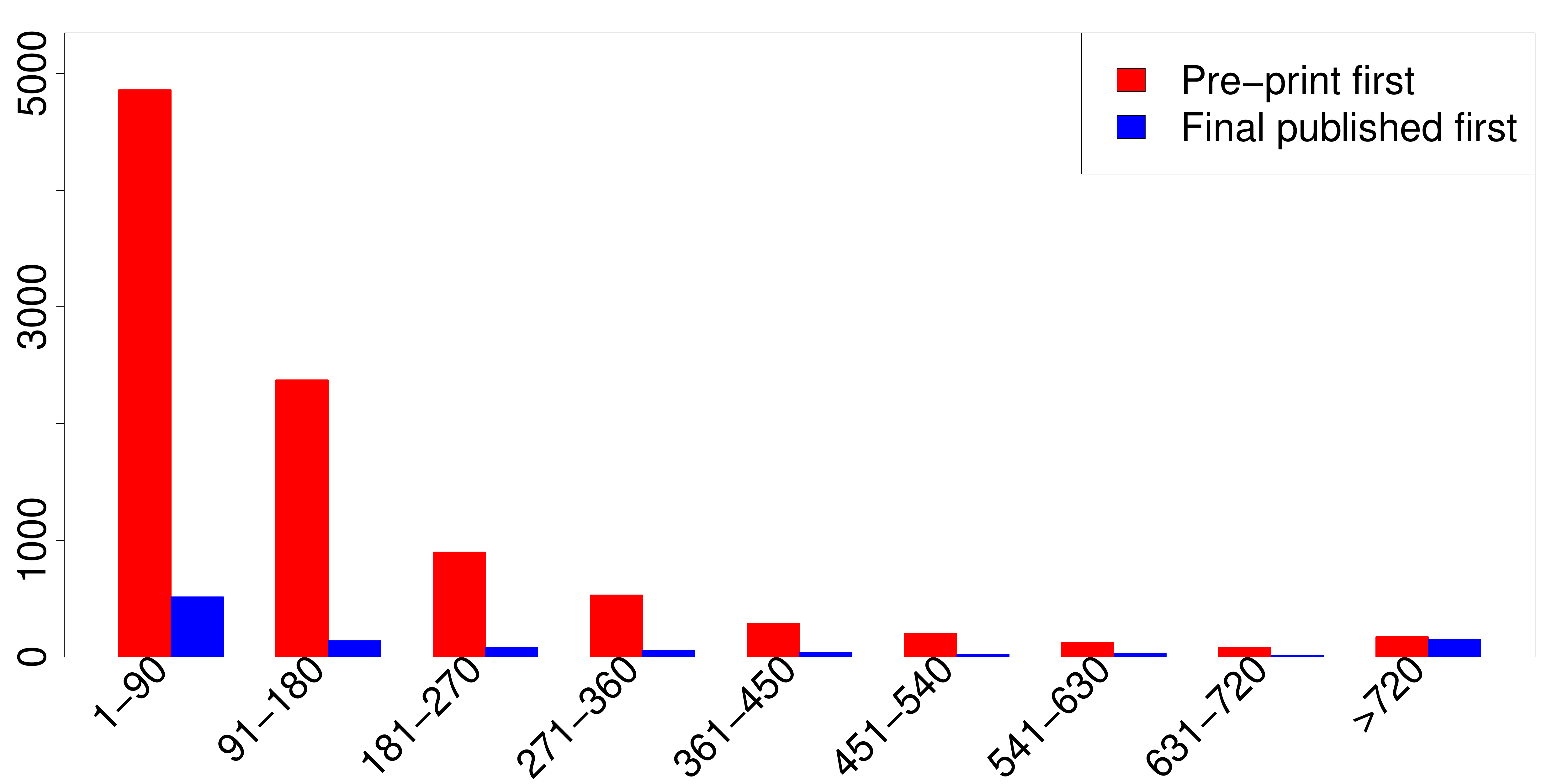}
\caption{Numbers of articles from the arXiv corpus first appearing in the specified venue, given the date 
of the \textbf{last} pre-print upload and the commercial publication date, binned by the number of days between them}
\label{fig:pub_dates}
\end{figure}

The above results provide strong indicators that there is hardly any noticeable difference between 
the pre-print version of a paper and its final published version. However, the results do not show
which version came first. In other words, consider the two possible scenarios:
\begin{enumerate}
\item Papers, after having gone through a rigorous peer review process, are published by a commercial 
publisher first and then, as a later step, uploaded to arXiv.org. In this case the results of our 
text comparisons described above would not be surprising, as the pre-print versions would merely be 
a mirror of the final published ones. There would be no apparent reason to deny publishers all credit 
for peer review, copyediting, and the resulting publication quality of the articles.
\item Papers are uploaded to arXiv.org first and later published by a commercial publisher. If this 
scenario is dominant, our comparison results would suggest that any changes in the text due to 
publisher-initiated copyediting are hardly noticeable.
\end{enumerate}
Figure \ref{fig:pub_dates} shows the order of appearance in arXiv.org versus commercial venues for 
all articles in our corpus, comparing the publication date of each article's final published version 
to the date of its latest upload to arXiv.org. Red bars indicate the amount of articles (absolute 
values on the y-axis) that were first upload to arXiv.org, and blue bars stand for articles published 
by a commercial publisher before they appeared in arXiv.org. Each pair of bars is binned into a time 
range, shown on the x-axis, that indicates how many days passed between the article's 
appearance in the indicated first venue and its appearance in the second venue.
Figure \ref{fig:pub_dates} shows clear evidence that the vast majority of our articles ($90\%$) were 
published in arXiv.org first. Therefore our argument for the second scenario from above holds. 
We can only speculate about the causes of certain time windows' prominence within the distribution, 
but it may be related to turn-around times of publishers between submission and eventual publication.
\section{Versions of Articles from the arXiv.org Corpus} \label{sec:article_versions}
About $35\%$ of all $1.2$ million papers in arXiv.org at time of writing have more than one version. 
A new version is created when, for example, an author makes a change to the article and re-submits 
it to arXiv.org. The evidence of Figure \ref{fig:pub_dates} shows that the majority of the latest 
versions in arXiv.org were uploaded prior to the publication of its final published version in a 
commercial venue. However, we were motivated to eliminate all doubt and hence decided to repeat 
our comparisons of the text contents of paper titles, abstracts, and body sections using the 
earliest versions of the articles from arXiv.org only. The underlying assumption is that those 
versions were uploaded to arXiv.org even earlier (if the authors uploaded more than one version) 
and hence are even less likely to exhibit changes due to copyediting by a commercial publisher. 
It follows, then, that if the comparisons of these earlier pre-print texts to their published 
versions show substantially greater divergences, then it is possible that more of these changes 
are the result of publisher-initiated copyediting. 
\begin{figure}[t!]
\center
\includegraphics[scale=0.3]{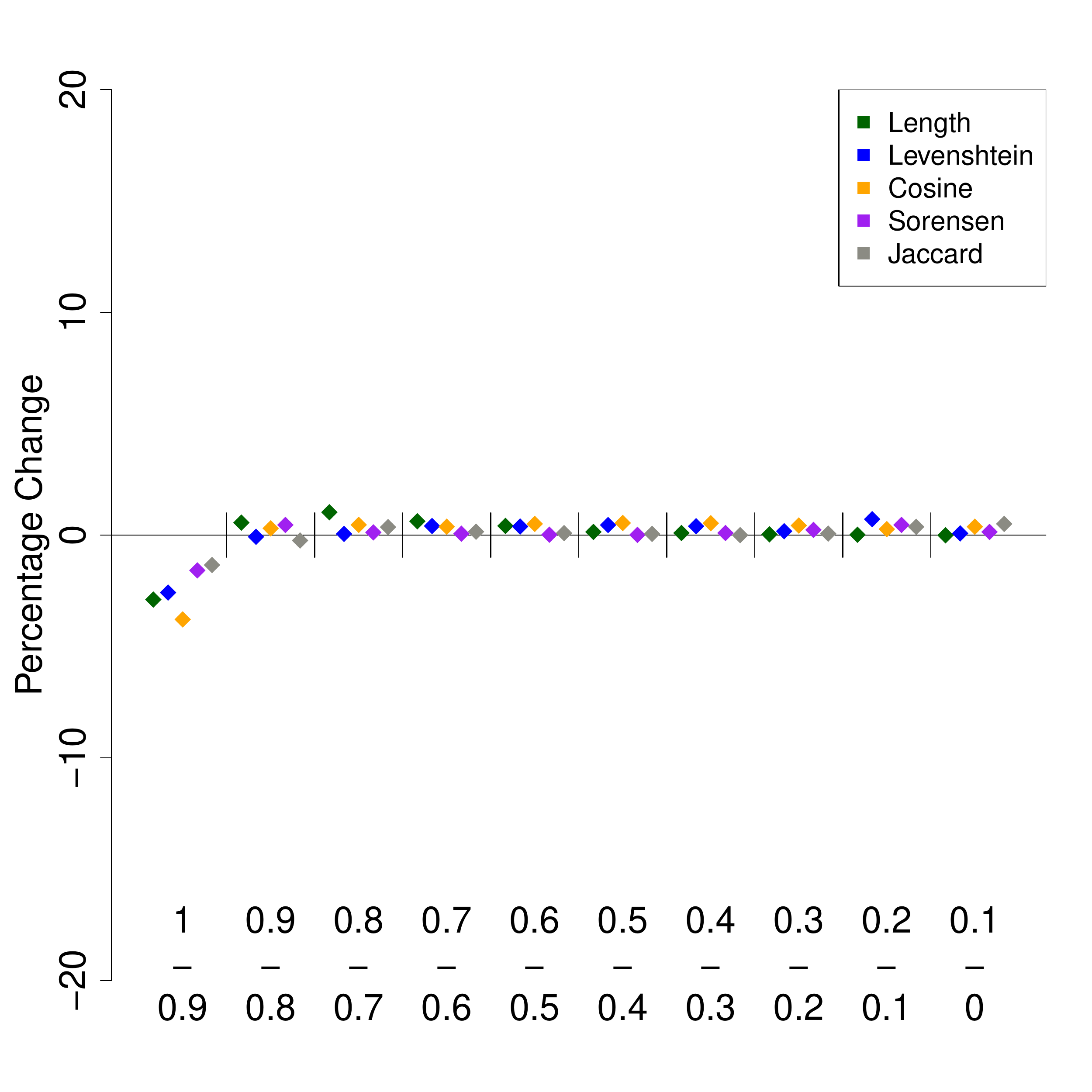}
\caption{Deltas resulting from the comparison of the title sections of the first uploaded pre-print to the final published paper,
contrasted with the comparison of the last uploaded pre-print to the final published paper.}
\label{fig:title_delta}
\end{figure}
\begin{figure}[t!]
\center
\includegraphics[scale=0.3]{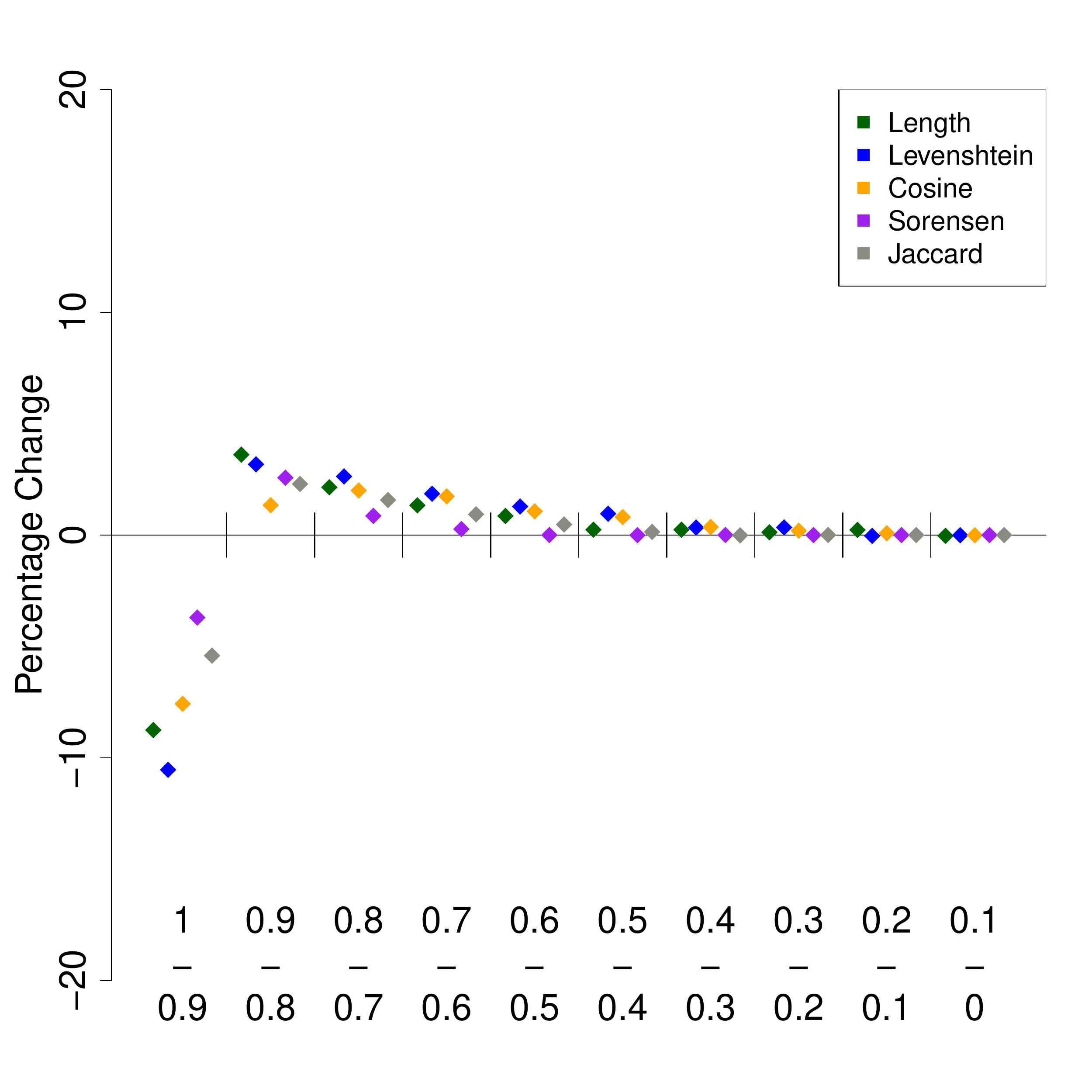}
\caption{Deltas resulting from the comparison of abstracts, as in Figure \ref{fig:title_delta}.}
\label{fig:abstract_delta}
\end{figure}
\begin{figure}[t!]
\center
\includegraphics[scale=0.3]{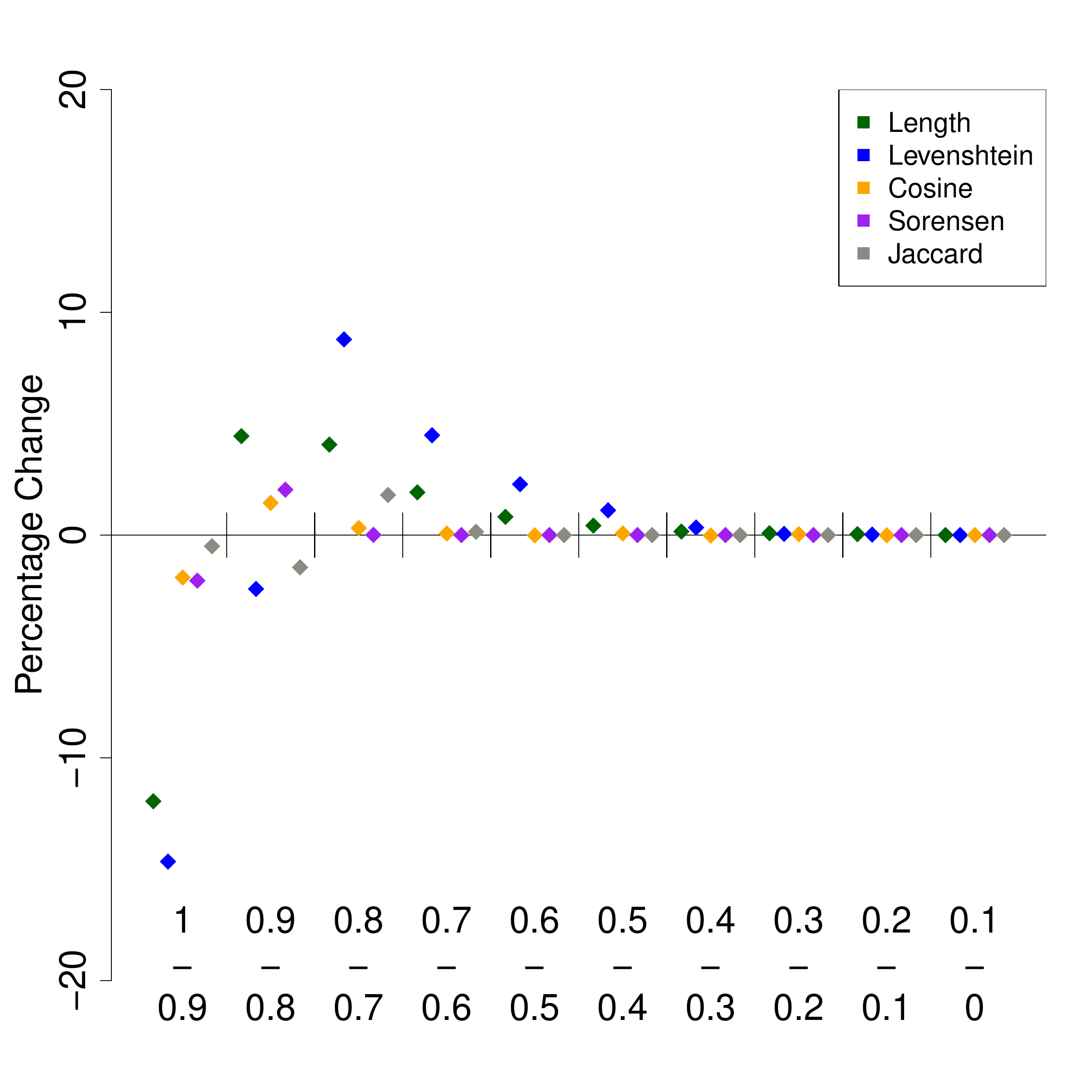}
\caption{Deltas resulting from the comparison of the body, as in Figure \ref{fig:title_delta}.}
\label{fig:body_delta}
\end{figure}

Our corpus of pre-print and final published papers matched by their DOIs and available via UCLA's 
journal subscriptions exhibits a higher ratio of papers with more than one version in arXiv.org 
than is found in the full set of articles available from arXiv.org. $58\%$ of the papers we 
compared had more than one version, $39\%$ had exactly two, and $13\%$ had exactly three versions; 
whereas only $35\%$ of all articles uploaded to arXiv.org have more than one version. We applied 
our five similarity measures (see Section \ref{subsec:txt_comp_meth}) to quantify the similarity
between the first versions of all articles and their final published versions. Rather than 
repeating the histograms of Figures \ref{fig:title_histo}, \ref{fig:abstract_histo}, 
and \ref{fig:body_histo}, we show the divergences from these histograms only. 

Figure \ref{fig:title_delta} depicts the deltas of the relative values of the title comparison. 
The colored dots represent the similarity measures applied, and as seen in previous figures, the five 
colors corresponding to the five similarity measures are grouped into bins. The values on the y-axis 
represent the delta between the relative numbers from the last and the first versions. 
For example, in Figure \ref{fig:title_histo}, which shows the title comparison numbers of the last 
pre-print versions, we see that $64.9\%$ of cosine values fall into the top bin. For the title comparison 
of first pre-print versions, only $61.1\%$ of cosine values fall into the top bin. We subtract the 
former value from the latter and arrive at a value of $-3.8$. This value is represented by the yellow dot 
(for the cosine similarity measure) in the leftmost bin of Figure \ref{fig:title_delta}.
Another example is the cosine value for the second bin. For the last versions, $5.6\%$ of values fall into
this bin, compared to $5.9\%$ for the first versions. Hence the yellow dot in Figure \ref{fig:title_delta} 
represents a value of $0.3$.
Therefore, if a delta value is negative, fewer comparisons of the first pre-print to the final published version
result in scores that fall into the corresponding bin for the last pre-print comparsion. Positive values, on the 
other hand, show that more comparison scores fall into a particular bin for the first rather than the last versions.
Figure \ref{fig:title_delta} shows small negative values for the top bin, positive numbers for the second and third
bins, and basically unchanged values for all other bins. This means the title comparison results are very similar 
between the last pre-print versions and the final published versions and the first pre-print versions and the
final published versions.
Figure \ref{fig:abstract_delta} presents the delta values for the abstract comparisons. The numbers are
fairly similar and show that the similarity of abstracts for first versions is slightly lower compared
to last pre-print versions. The deltas in the second, third, fourth, and all following bins, however, are 
all positive, which indicate that the differences are not substantial. 
The numbers for the body comparison are shown in Figure \ref{fig:body_delta}. 
We can observe that the length ratio and the Levenshtein scores for the top bin are $11.9$ and $14.7$ points
lower. However, the cosine scores are fairly similar, indicating that the semantic-level similarity is still
high.

These results confirm our initial assessment that very little difference can be found between 
pre-print articles and their final published versions. Even more so, these findings strengthen our 
argument as they show that the difference between the earliest possible pre-print version and the 
final published one seems insignificant, given the similarity measures we applied to our corpus.
\begin{figure}[t!]
\center
\includegraphics[scale=0.17]{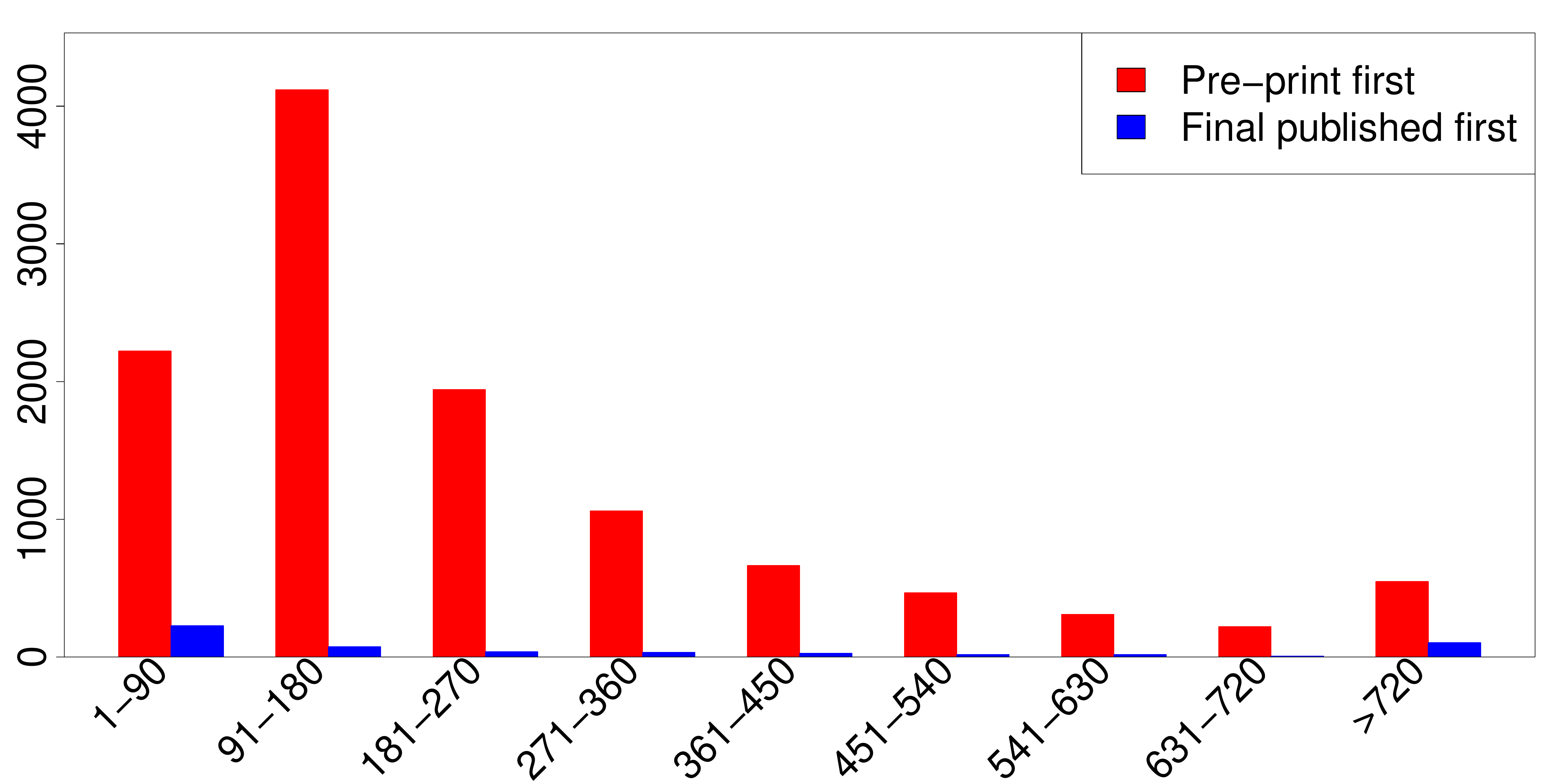}
\caption{Numbers of articles in the arXiv corpus first appearing in the specified venue, given the date of the 
\textbf{first} pre-print upload and the commercial publication date, binned by the number of days between them}
\label{fig:pub_dates_versions}
\end{figure}
\begin{figure*}[t!]
\center
\includegraphics[scale=0.5]{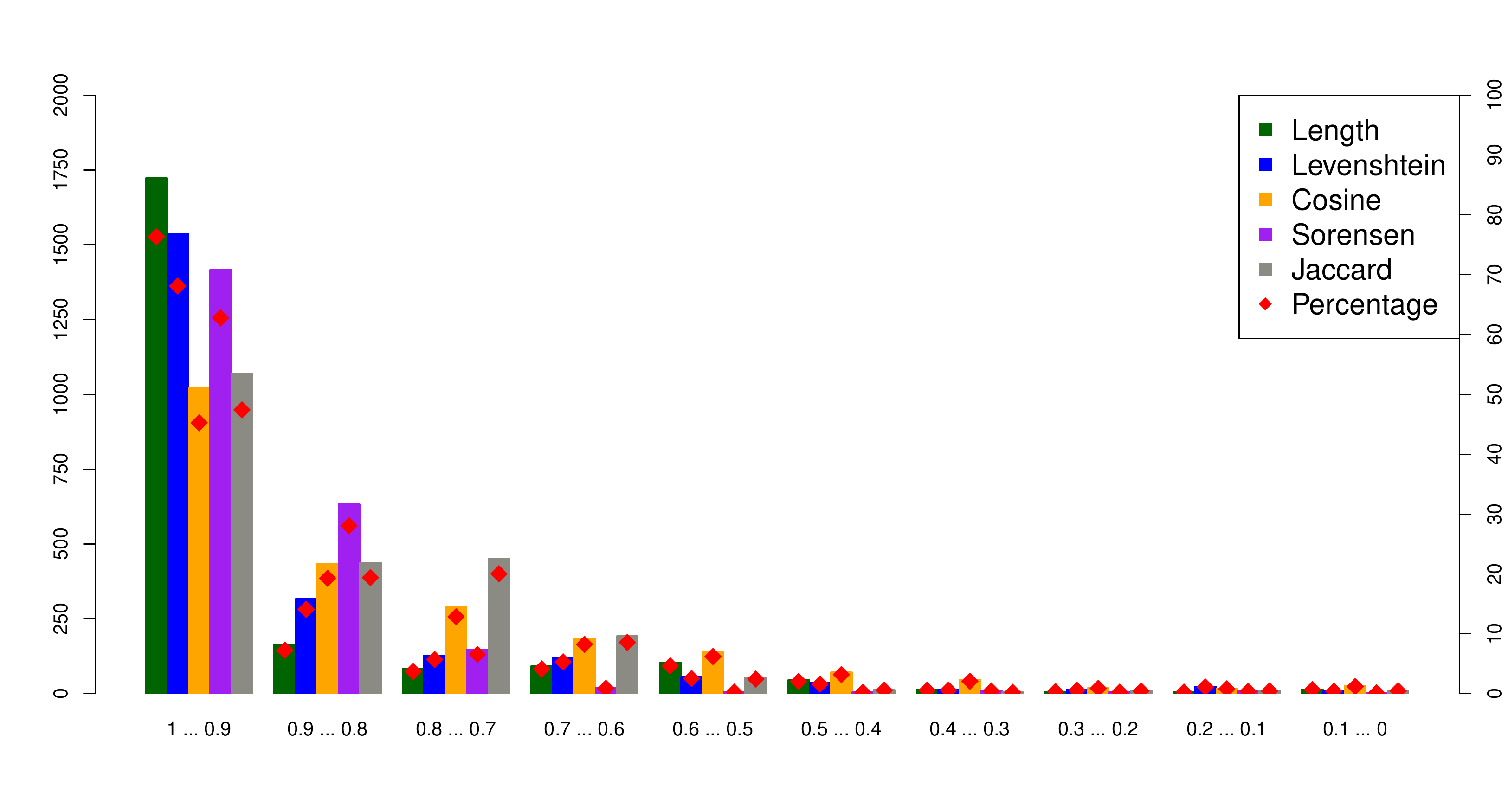}
\caption{bioRxiv corpus - comparison results for titles}
\label{fig:biorxiv_title_histo}
\end{figure*}
\subsection{Publication Dates of Versions}
The scenarios discussed in Section \ref{subsec:pub_dates} with respect to the question of whether 
an article was uploaded to arXiv before it appeared in a commercial venue are valid for this 
comparison as well. 
Figure \ref{fig:pub_dates_versions} mirrors the concept of Figure \ref{fig:pub_dates} and 
shows the number of earliest pre-print versions uploaded to arXiv.org first in red
and the final published versions appearing first represented by the blue bars.
As expected, the amount of pre-print versions published first increased and now stands at $95\%$ 
as shown in Figure \ref{fig:pub_dates_versions} (compared to $90\%$ shown in 
Figure \ref{fig:pub_dates}). Our argument for the second scenario described above is therefore 
strongly supported when considering the earliest uploaded versions of pre-prints. 
\section{bioRxiv Corpus\\Experiment Results} \label{sec:biorxiv_results}
We were curious whether the generally neglibible differences detected between pre-print and final 
published versions of articles from arXiv.org also would be prevalent among papers from a different 
scientific domain --- thereby suggesting whether or not further replicative studies might 
find this phenomenon to be general across STM fields. As discussed above, the life sciences 
pre-print repository bioRxiv.org, which was explicitly modeled on arXiv.org, proved to be the most 
readily available source of pre-prints for this follow-on work, and we were able to accumulate a 
sufficient corpus of matching published versions for comparison. We therefore calculated similarity 
values for each pair of corresponding papers, again comparing titles, abstracts, and body sections, 
producing visualizations very similar in nature to the previously seen figures. The results of these 
comparisons, as presented in further detail below, shared the same overall characteristics of
those from the arXiv data set, although they exhibited some differences as well. We attribute
these differences primarily to divergent disciplinary practices between physics and biology (and
their related fields) with respect to the degrees of formatting applied to pre-print and published 
articles.

\begin{figure*}[t!]
\center
\includegraphics[scale=0.5]{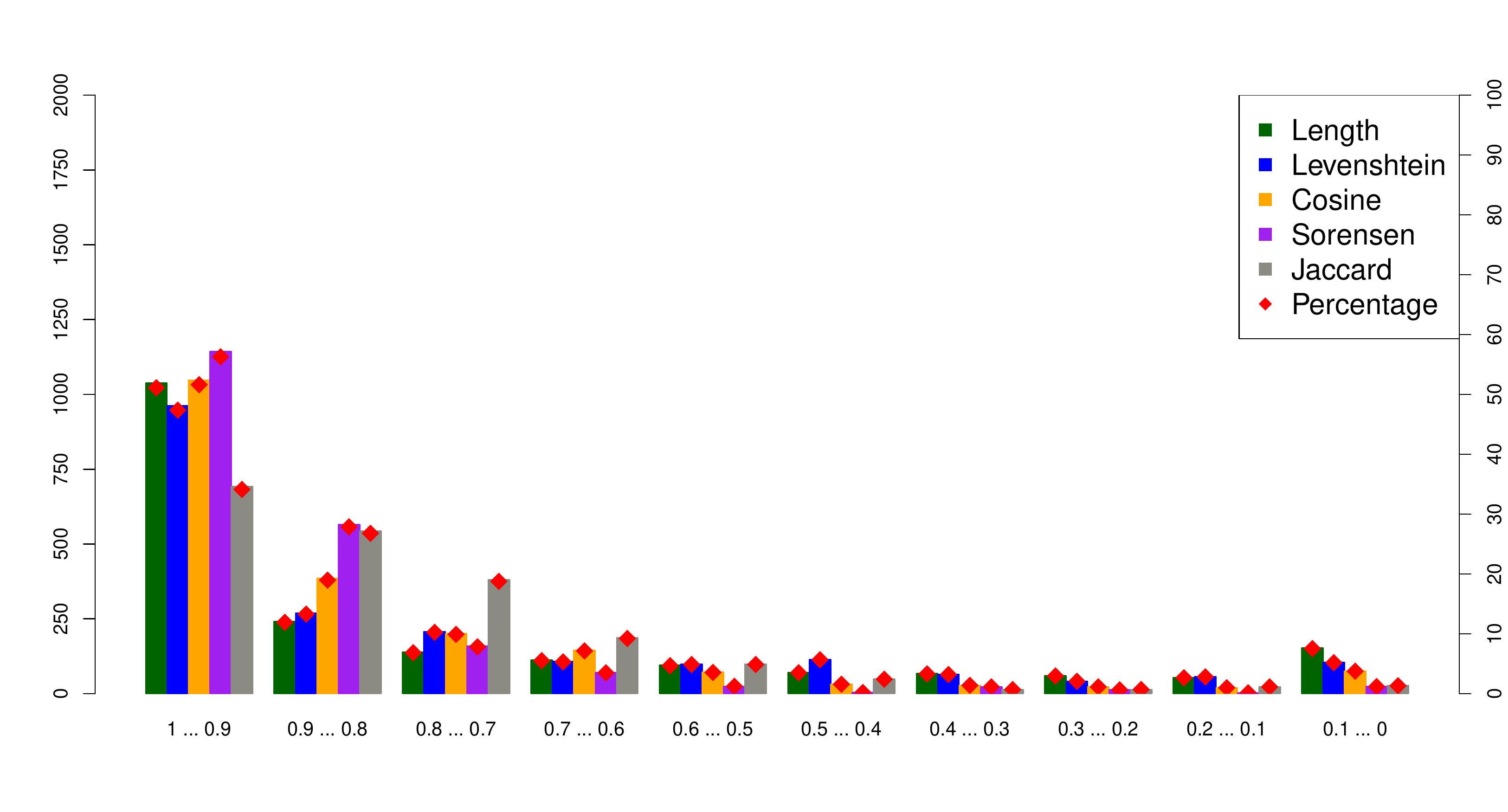}
\caption{bioRxiv corpus - comparison results for abstracts}
\label{fig:biorxiv_abstract_histo}
\end{figure*}
\subsection{Title Analysis}
Figure \ref{fig:biorxiv_title_histo} shows the scores of all five similarity measures for the title comparison,
which is quite similar in concept to Figure \ref{fig:title_histo} for the arXiv corpus. We can again observe the 
dominance of the top bin, with more than $76\%$ of length scores, $68\%$ of Levenshtein scores, and $62\%$ of 
S\o rensen scores falling into this bin. Just less than half of all cosine and Jaccard scores also range between 
$1$ and $0.9$. The vast majority of remaining scores fall into the second and third bin, and only cosine and Jaccard 
see around $8\%$ of scores in the fourth bin from the left. These results from the bioRxiv corpus confirm our earlier 
findings from the arXiv corpus that titles of scholarly articles rarely change noticeably between the pre-print and 
the final published version.
\subsection{Abstract Analysis}
The bars shown in Figure \ref{fig:biorxiv_abstract_histo} represent the comparison scores for abstracts from the 
bioRxiv corpus. The graph is similarly dominated by the top bin, but compared to abstracts from the arXiv corpus 
(Figure \ref{fig:abstract_histo}) the numbers are more evenly distributed. Around half of all length, Levenshtein, 
and cosine values fall into the top bin, along with $34\%$ of Jaccard scores. Given that the majority of the 
remaining scores fall into the second, third, and fourth bins, we can confidently say that all similarity measures 
score very high for bioRxiv abstracts. We do note, however, a small percentage of scores falling into the last bin 
with values between $0.1$ and $0$, indicating that some astracts have significantly changed in length and even in 
terms of semantic resemblance, which is indicated by the cosine score. 
\subsection{Body Analysis} 
\begin{figure*}[t!]
\center
\includegraphics[scale=0.5]{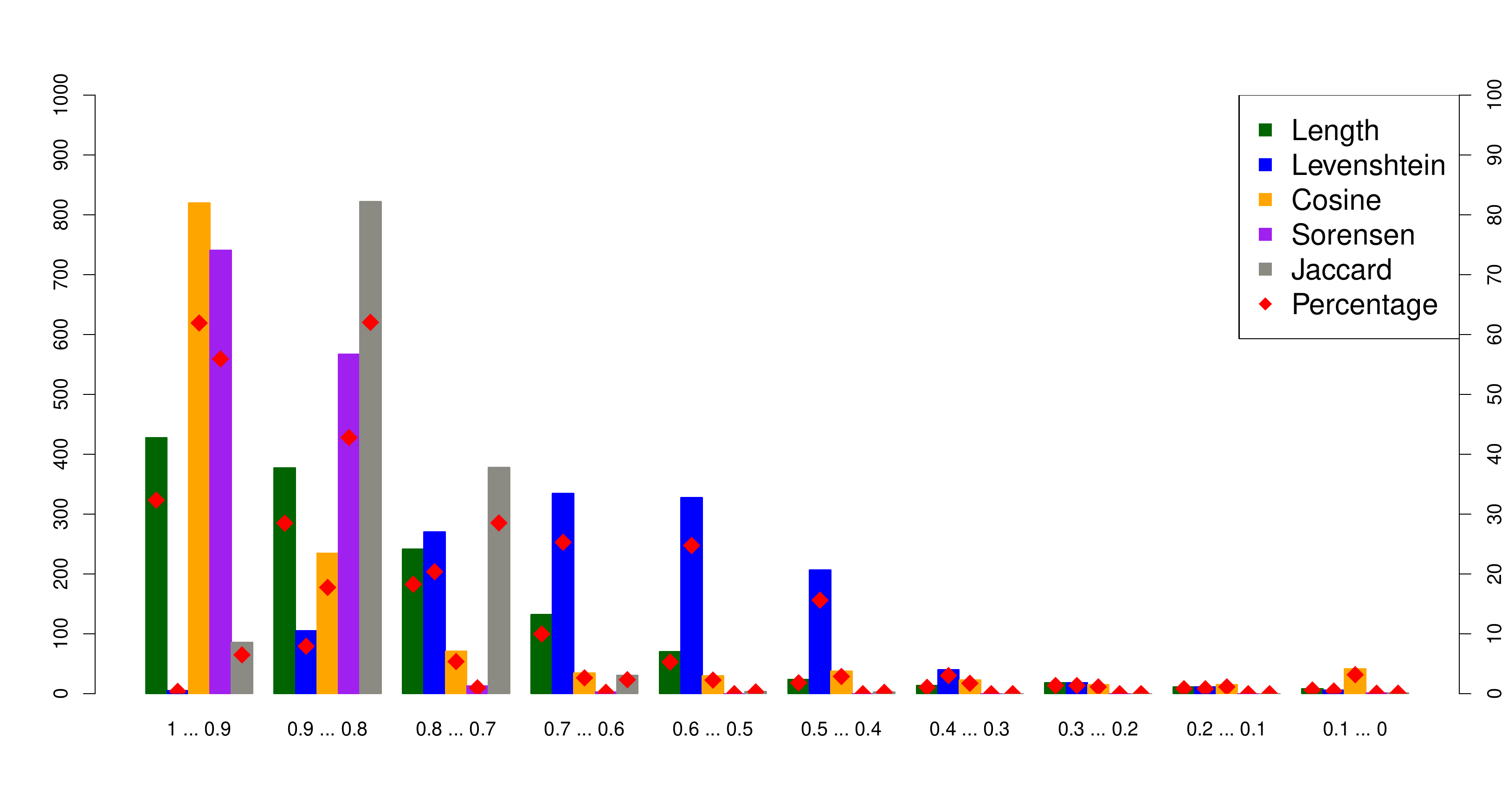}
\caption{bioRxiv corpus - comparison results for body sections}
\label{fig:biorxiv_body_histo}
\end{figure*}
The results of the comparison between body sections are shown in Figure \ref{fig:biorxiv_body_histo}. It is 
astonishing to see the very high percentage of cosine scores, $62\%$, in the top bin. Furthermore, $85\%$ of all 
cosine scores fall into the top three bins. The top two bins hold almost all of the S\o rensen scores and the top 
three hold almost all Jaccard scores --- both are set-based similarity scores, yet again they indicate a very high 
level of similarity of the compared body sections. The Levenshtein scores are, especially compared to the arXiv 
corpus (Figure \ref{fig:body_histo}), slightly shifted to the right. While the vast majority of values are above 
$0.4$, almost no scores fall into the top bin. 
This observation can likely be attributed to the nature of the bioRxiv corpus, where we have observed pre-print versions 
adhering to simple templates that were changed for the final published version. The Levenshtein similarity measure 
is most sensitive to these sorts of textual differences.
For the body comparison, we again see a small number of cosine scores ($3.2\%$) falling into the last bin, which 
indicates dramatically different content between the compared bodies of text.
Regardless, the overall very high scores, especially the semantic and set-based scores, provide additional evidence 
that bioRxiv pre-prints and their final published versions do not exhibit many differentiating features, neither on 
the editorial nor on the semantic level. 
\begin{figure}[th!]
\center
\includegraphics[scale=0.17]{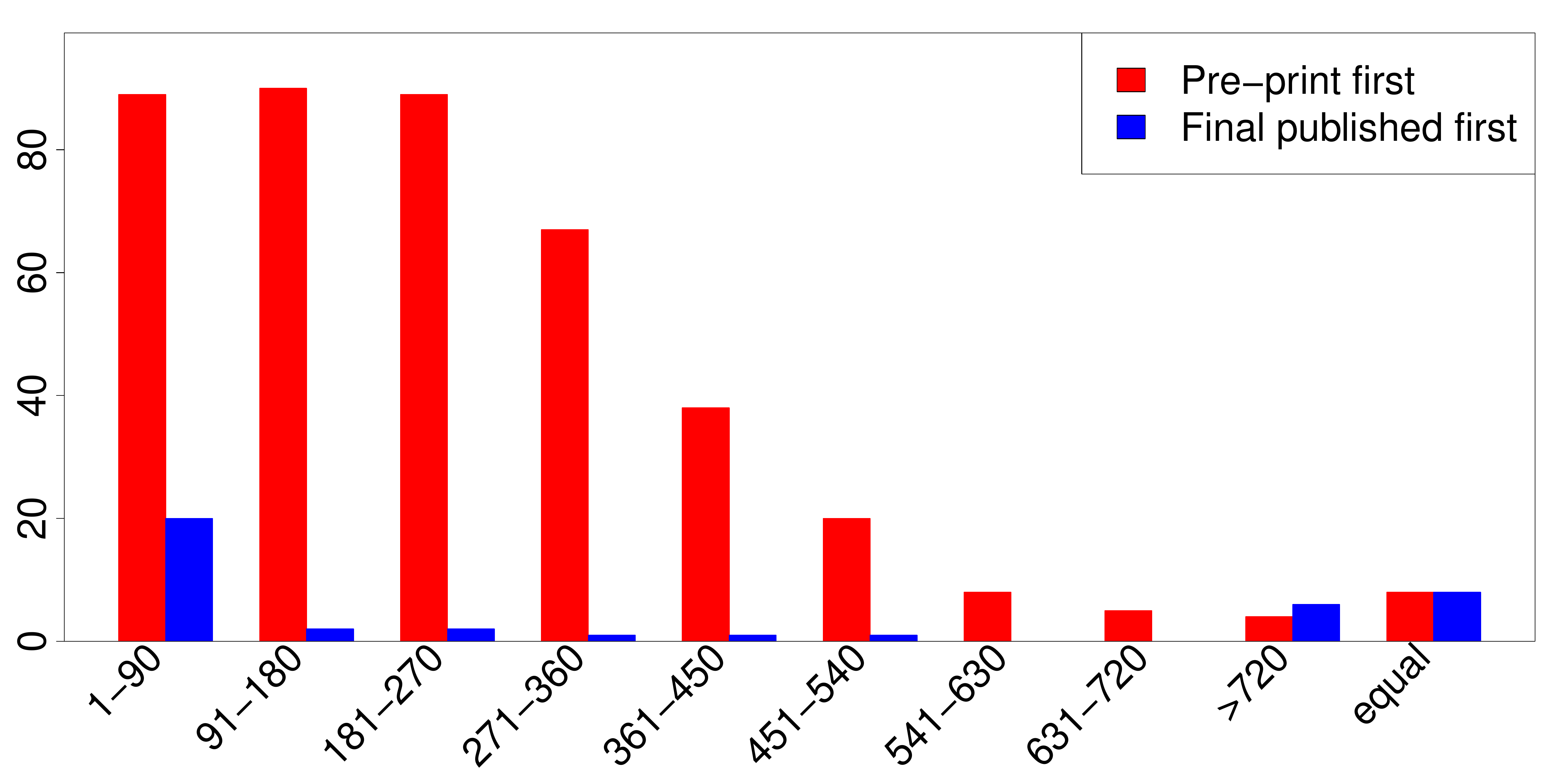}
\caption{Numbers of articles in the bioRxiv corpus first appearing in the specified venue, given the date of the 
\textbf{last} pre-print upload and the commercial publication date, binned by the number of days between them}
\label{fig:biorxiv_pub_dates}
\end{figure}
\subsection{Publication Dates}
We also analyzed the scenarios discussed in Section \ref{subsec:pub_dates} with respect to the question of whether or
not an article appears in the pre-print repository before it is published in a journal. The ratio of bioRxiv articles 
appearing first as pre-prints (red bars) versus appearing first as final published versions (blue bars) is shown in 
Figure \ref{fig:biorxiv_pub_dates}. Similar to Figure \ref{fig:pub_dates}, the bars are binned into time ranges 
showing the number of days since first publication in the respective venue. Figure \ref{fig:biorxiv_pub_dates} shows 
that $91\%$ of bioRxiv articles were published as pre-print first. This number is similar to articles from the arXiv 
corpus, and is also not surprising given bioRxiv's stated purpose of archiving non-refereed pre-prints. Perhaps as
a consequence of this preference for pre-submission versions, we do not see the same dominance of the $1-90$ days
time slot that was observed with arXiv articles. In fact, $82\%$ of bioRxiv articles that appear as pre-prints first 
are posted on bioRxiv anywhere from $1$ to $360$ days before their final published counterpart is published by a 
commercial publisher. $2\%$ of articles had identical dates of appearance in bioRxiv and their final published venue.
\section{Versions of Articles from the bioRxiv Corpus}
bioRxiv articles can have multiple versions, which is another resemblance to the arXiv repository that served as the 
model for its creators. A new version is generated when an author resubmits a modified verion of a paper or 
edits the article-specific metadata. $933$ ($40\%$) of the $2,332$ papers in our pre-print corpus from bioRxiv had more 
than one version (compared to $58\%$ in the arXiv corpus). 
Prompted by motivations that were quite similar to those described in Section \ref{sec:article_versions}, we conducted 
the same similarity experiments we applied to papers from the arXiv corpus with multiple pre-print versions (see 
Section \ref{subsec:txt_comp_meth}) but considered only the first versions of papers as uploaded to bioRxiv and 
compared them to their final published counterparts. Naturally, these versions were uploaded prior to the last versions 
in the bioRxiv corpus that we considered in the previous experiments (detailed in Section \ref{sec:biorxiv_results}) 
and hence, intuitively, should show fewer indicators of copyediting by commercial editors. 

\begin{figure}[t!]
\center
\includegraphics[scale=0.3]{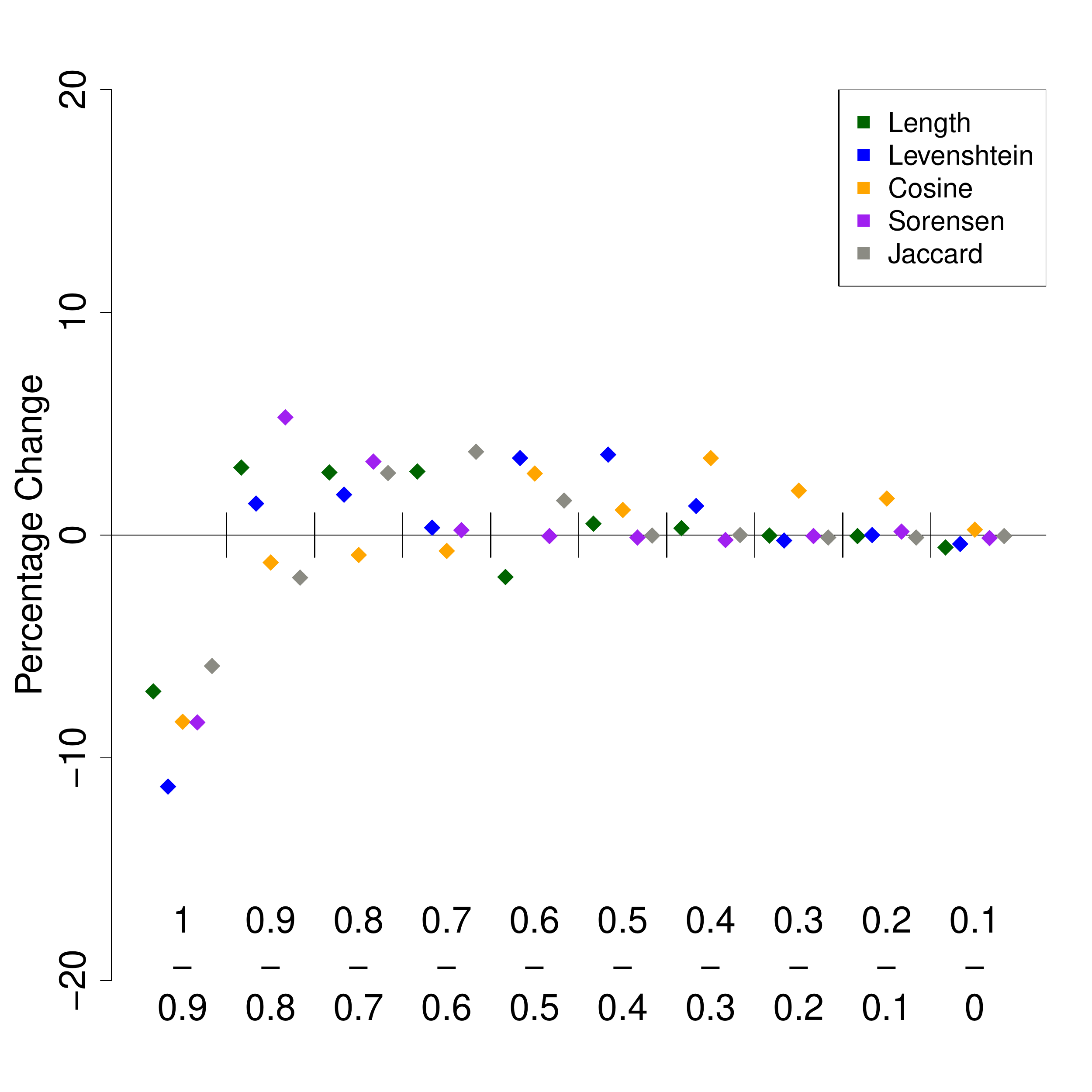}
\caption{bioRxiv corpus - deltas resulting from the comparison of the title sections of the first uploaded pre-print to the 
final published paper, contrasted with the comparison of the last uploaded pre-print to the final published paper.}
\label{fig:biorxiv_title_delta}
\end{figure}

\begin{figure}[ht!]
\center
\includegraphics[scale=0.3]{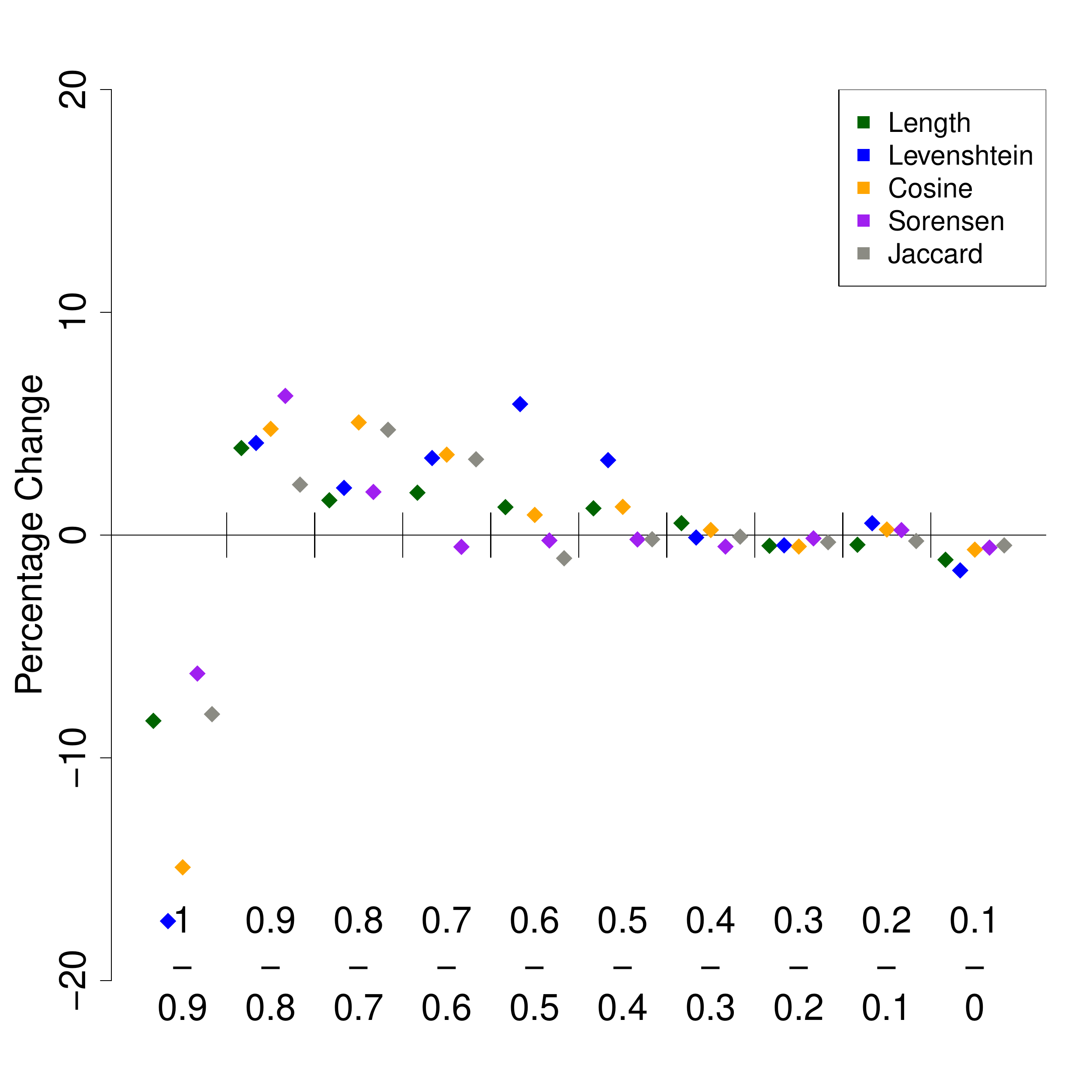}
\caption{bioRxiv corpus - deltas resulting from the comparison of abstracts, as in Figure \ref{fig:biorxiv_title_delta}.}
\label{fig:biorxiv_abstract_delta}
\end{figure}

\begin{figure}[ht!]
\center
\includegraphics[scale=0.3]{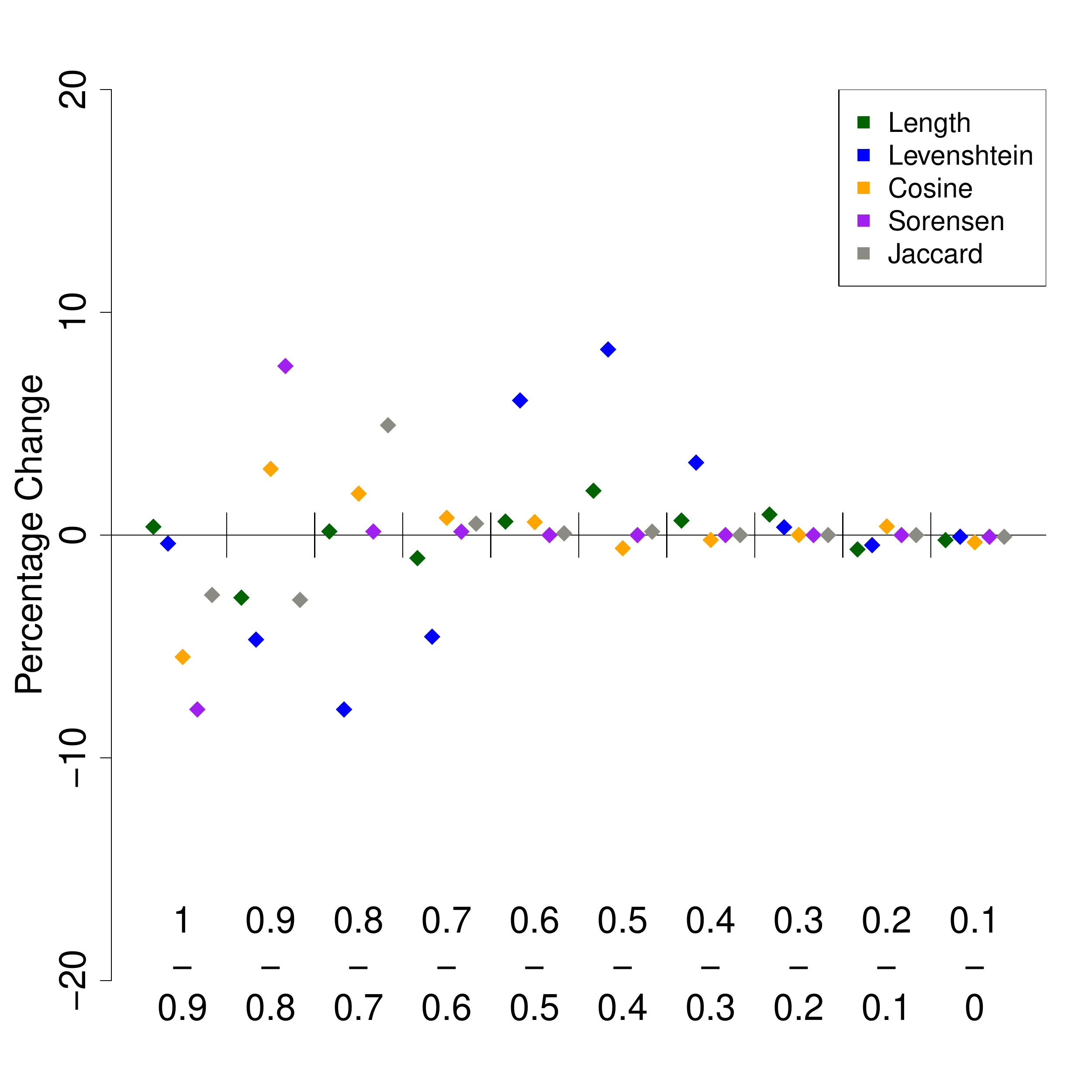}
\caption{bioRxiv corpus - deltas resulting from the comparison of the papers' body sections, as in Figure \ref{fig:biorxiv_title_delta}.}
\label{fig:biorxiv_body_delta}
\end{figure}

The visualizations of the differences between these comparisons are very similar to the previously seen 
Figures \ref{fig:title_delta}, \ref{fig:abstract_delta}, and \ref{fig:body_delta} for the arXiv corpus.
Figure \ref{fig:biorxiv_title_delta} shows the differences in relative scores for the title comparisons. We can observe
that for all similarity measures, fewer scores fall into the top bin. However, the majority of delta values in the following 
bins are positive, which means the similarity scores shift from the top bin to the second, third, and fourth bins. An expection 
is the delta for the cosine score, which is negative for the first four bins and just then turns positive. This
indicates that there indeed are more semantic changes in the titles of the first bioRxiv versions compared to their final
published versions. We do not see this pattern for the first arXiv versions (Figure \ref{fig:title_delta}).
Figure \ref{fig:biorxiv_abstract_delta} displays the changes in relative scores for the abstract comparison. We see a similar
pattern with fewers scores of all similarity measures falling into the top bin. However, unlike the score changes for titles,
we see here that all similarity scores have positive deltas in the following bins. In particular, a fraction of length, cosine, 
S\o rensen, and Jaccard values that fell into the top bin for the last versions are now distributed over the second, third, 
and fourth bins. The Levenshtein values drop the most from the top bin and seem to frequently fall into bins representing 
similarity scores between $0.6$ and $0.5$ as well as $0.5$ and $0.4$. 
The changes of relative values for the body comparison are shown in Figure \ref{fig:biorxiv_body_delta}.
The patterns are very similar to what we have observed for titles and abstracts. The delta for the Levenshtein scores stands 
out, however. Figure \ref{fig:biorxiv_body_histo} shows almost no Levenshtein scores in the top bin and 
Figure \ref{fig:biorxiv_body_delta} confirms that there is almost no change to this bin. The delta of scores in the three
consecutive bins is negative before turning positive in the fifth, sixth, and seventh bin. This means that the pattern shown in
Figure \ref{fig:biorxiv_body_histo} for the Levenshtein score is amplified and the scores are increasingly falling into bins
that represent lower scores. It is plausible that the explanation cited above involving the authors' use of rudimentary
article templates also applies here.
\subsection{Publication Dates of Versions}
\begin{figure}[h!]
\center
\includegraphics[scale=0.17]{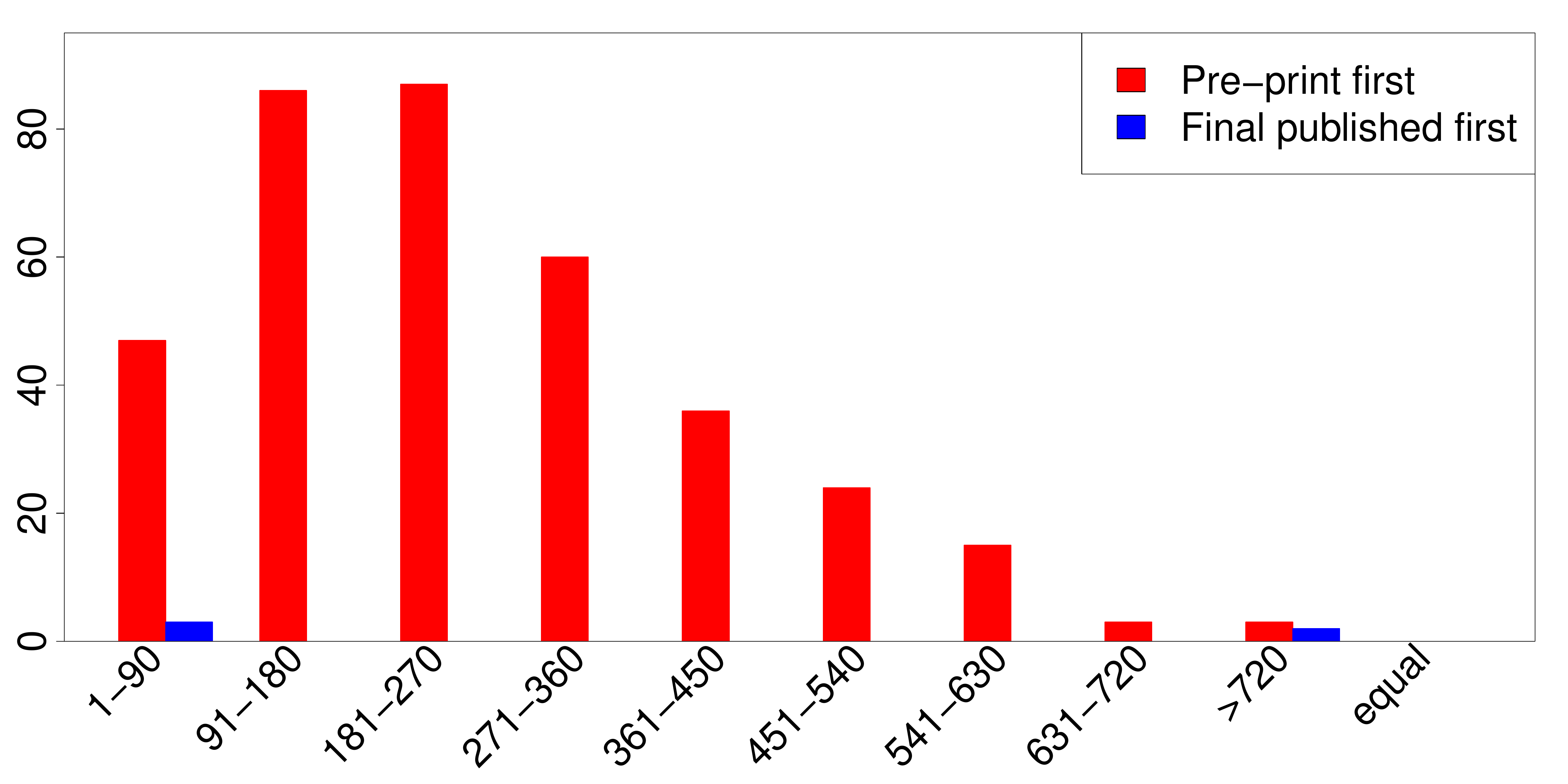}
\caption{Numbers of articles in the bioRxiv corpus first appearing in the specified venue, given the date of 
the \textbf{first} pre-print upload and the commercial publication date, binned by the number of days between them.}
\label{fig:biorxiv_pub_dates_versions}
\end{figure}

As described for the arXiv corpus in Section \ref{sec:article_versions}, we also investigated the publication dates
of the first versions of papers uploaded to bioRxiv.
Figure \ref{fig:biorxiv_pub_dates_versions} shows the ratios of first versions of papers first appering in bioRxiv to
those published by a commercial publisher first. Not unlike the ratios for last versions in bioRxiv as shown in 
Figure \ref{fig:biorxiv_pub_dates}, we can observe a clear dominance of papers appearing in bioRxiv first. $99\%$ of
papers fall into this category (compared to $91\%$ for last versions of bioRxiv papers). 
As these papers are first versions in bioRxiv, it is not surprising that the time difference between their pre-print
upload date and the date they were finally published by a commercial publisher increases relative to the last version
uploaded to bioRxiv. We made the same observation for the arXiv corpus in Figure \ref{fig:pub_dates_versions}, 
compared to Figure \ref{fig:pub_dates}.
\section{Discussion and Future Work}
The results outlined in this paper are from a preliminary study on the similarity of pre-print 
articles to their final published counterparts. There are many areas where this study could be 
improved and enhanced.
Expanding this line of experiments to other domains such as the humanities, social sciences, and 
economics might return different results, as the review and editorial practices in other 
disciplines can vary considerably. 

The matching of a pre-print version of an article to its final published version was done by means 
of the article's DOI. While this is an obvious choice for a paper identifier, by only relying on 
DOIs we very likely missed out on other matching articles.
Note also that we could only match articles that we have access to via UCLA Library's serial 
subscriptions or via open access publications. It might be worth expanding the matching process 
to a collaborating organization with ideally complementary subscriptions to maximize access to full 
text articles.

One typical article section we have not analyzed as part of this research is the references section. 
Given publishers' claims of adding value to this section of a scholarly article, we are motivated 
to see whether we can detect any significant changes between pre-prints and final published versions 
there. Similarly, we have not thoroughly investigated changes in the author sections. We anticipate 
author movement, such as authors being added, being removed, and having their rank in the list of 
authors changed --- although changes in author order due to publishers' name alphabetization 
policies must be considered as well. Initial experiments in this domain have proven difficult to 
interpret, as author names are provided in varying formats and normalization is not trivial. 

Another angle of future work is to investigate the correlation between pre-prints and final 
published versions' degree of similarity and measured usage statistics such as download numbers 
and the articles' impact factor values. When arguing that the differences between pre-print articles 
and their final published versions are insignificant, factoring in usage statistics and 
``authority values'' can further inform decisions about investments in serial subscriptions.
\section{Conclusions}
This study is motivated by academic publishers' claims of the value they add to scholarly articles 
by copyediting and making further enhancements to the text. We present results from our preliminary 
study to investigate the textual similarity of scholarly pre-prints and their final published 
counterparts. 
We generate two different corpora from the popular pre-print services arXiv and bioRxiv and match
their papers to the corresponding versions as published by commercial publishers. We use standard
text extraction methods to compare individual sections of papers such as the title, abstract, and
the body. For the text comparison, we apply five different similarity measures and analyze their 
results. 

We have shown that, within the boundaries of the arXiv corpus, there are no significant differences 
in aggregate between pre-prints and their corresponding final published versions. The picture for
the bioRxiv corpus is very similar, but we do see a slightly larger divergence between pre-print
and final published paper versions in this case, suggesting that varying disciplinary practices 
regarding formatting and copyediting can and do influence the degree of detected similarity 
between pre-print and final published articles.
In addition, we have shown for both corpora that the vast majority of pre-prints ($90\%$ and $91\%$,
respectively) are published by the open access pre-print services first and later by a commercial 
publisher. If we consider the first ever uploaded pre-print versions, these numbers increase to
$95\%$ and $99\%$, respectively. 

Given the fact of flat or even shrinking library, college, and university budgets, our 
findings provide empirical indicators that should inform discussions about commercial publishers' 
value propositions in scholarly communication and have the potential to influence higher education 
and academic and research libraries' economic decisions regarding access to scholarly publications.
%
%
%\bibliographystyle{abbrv}
%\bibliographystyle{spbasic}
%\bibliographystyle{spmpsci}
%\bibliography{references}

\begin{thebibliography}{10}
\providecommand{\url}[1]{{#1}}
\providecommand{\urlprefix}{URL }
\expandafter\ifx\csname urlstyle\endcsname\relax
  \providecommand{\doi}[1]{DOI~\discretionary{}{}{}#1}\else
  \providecommand{\doi}{DOI~\discretionary{}{}{}\begingroup
  \urlstyle{rm}\Url}\fi

\bibitem{bjork:megajournals}
Bj{\"o}rk, B.C.: Have the ``mega-journals'' reached the limits to growth?
\newblock PeerJ \textbf{3}, e981 (2015)

\bibitem{bjork:openaccess}
Bj{\"o}rk, B.C., Welling, P., Laakso, M., Majlender, P., Hedlund, T.,
  Gu{\dh}nason, G.: Open access to the scientific journal literature: Situation
  2009.
\newblock PLoS ONE \textbf{5(6)}, e11,273 (2009)

\bibitem{bornmann:2015growth}
Bornmann, L., Mutz, R.: Growth rates of modern science: A bibliometric analysis
  based on the number of publications and cited references.
\newblock Journal of the Association for Information Science and Technology
  (2015)

\bibitem{jaccard:index}
Jaccard, P.: Etude comparative de la distribution florale dans une portion des
  Alpes et du Jura.
\newblock Impr. Corbaz (1901)

\bibitem{jamali:openaccess}
Jamali, H.R., Nabavi, M.: {Open access and sources of full-text articles in
  Google Scholar in different subject fields}.
\newblock Scientometrics \textbf{105}(3), 1635--1651 (2015)

\bibitem{levenshtein:edit_distance}
{Levenshtein}, V.I.: {Binary Codes Capable of Correcting Deletions, Insertions
  and Reversals}.
\newblock Soviet Physics Doklady \textbf{10}(8), 707--710 (1966)

\bibitem{mabe:ermh}
Mabe, M.: {(Electronic) Journal Publishing}.
\newblock The E-Resource Management Handbook  (2006)

\bibitem{budget:fy14}
{Office of Management and Budget (U.S.)}: Fiscal Year 2014 Analytical
  Perspectives: Budget of the U.S. Government.
\newblock Office of Management and Budget (2013)

\bibitem{pang:introdm}
Pang-Ning, T., Steinbach, M., Kumar, V.: {Introduction to Data Mining}.
\newblock Pearson Addison Wesley (2006)

\bibitem{porter:algo}
Porter, M.F.: An {A}lgorithm for {S}uffix {S}tripping.
\newblock Electronic Library and Information Systems \textbf{14}(3), 130--137
  (1980)

\bibitem{sorensen:index}
S{\o}rensen, T.: {A Method of Establishing Groups of Equal Amplitude in Plant
  Sociology Based on Similarity of Species and its Application to Analyses of
  the Vegetation on Danish Commons}.
\newblock Biol. Skr. \textbf{5}, 1--34 (1948)

\bibitem{ucla:accountability}
{University of California}: {Accountability Report 2015}.
\newblock
  \url{http://accountability.universityofcalifornia.edu/2015/chapters/chapter-9.html}

\bibitem{ware:stmreport}
Ware, M., Wabe, M.: {The STM Report - An Overview of Scientific and Scholarly
  Journal Publishing}.
\newblock International Association of Scientific, Technical and Medical
  Publishers (2015).
\newblock
  \urlprefix\url{http://www.stm-assoc.org/2015_02_20_STM_Report_2015.pdf}

\end{thebibliography}
%

%
%
%
\end{document}